\documentclass[final,5p,times,twocolumn]{elsarticle}

\usepackage{array}
\usepackage{longtable}
\usepackage{hyperref}
\usepackage{pgfplots}
\pgfplotsset{compat=1.18} 
\usepackage{multirow}
\usepackage{pdflscape}
\usepackage{graphicx}
\usepackage{geometry}
\usepackage{amssymb}
\usepackage{amsmath}
\usepackage{forest}
\usepackage{pgf-pie}
\usepackage{wrapfig}

\PassOptionsToPackage{round,numbers,sort&compress}{natbib}
\usepackage{natbib}

\begin{document}
\begin{frontmatter}

\title{Enhancing Research Methodology and Academic Publishing: A Structured Framework for Quality and Integrity}

\author[first]{Md. Jalil Piran}
\affiliation[first]{Department of Computer Science and Eingeering, Sejong University, Seoul 05006, South Korea, (email: piran@sejong.ac.kr)} 
\author[second]{Nguyen H. Tran}
\affiliation[second]{
The University of Sydney, School of Computer Science, 
1 Cleveland Street,, Darlington, NSW 2008, Australia,, email: (nguyen.tran@sydney.edu.au)}
\begin{abstract}
Following a brief introduction to research, research processes, research types, papers, reviews, and evaluations, this paper presents a structured framework for addressing inconsistencies in research methodology, technical writing, quality assessment, and publication standards across academic disciplines. Using a four-dimensional evaluation model that focuses on 1) technical content, 2) structural coherence, 3) writing precision, and 4) ethical integrity, this framework not only standardizes review and publication processes but also serves as a practical guide for authors in preparing high-quality manuscripts. Each of these four dimensions cannot be compromised for the sake of another. Following that, we discuss the components of a research paper adhering to the four-dimensional evaluation model in detail by providing guidelines and principles. By aligning manuscripts with journal standards, reducing review bias, and enhancing transparency, the framework contributes to more reliable and reproducible research results. Moreover, by strengthening cross-disciplinary credibility, improving publication consistency, and fostering public trust in academic literature, this initiative is expected to positively influence both research quality and scholarly publishing's reputation.

\end{abstract}

\begin{keyword}
Peer Review Process,
Research Quality Assessment,
Technical Evaluation,
Academic Publishing Standards,
Manuscript Structure,
Writing Quality,
Ethics Publishing,

\end{keyword}

\end{frontmatter}

\section{Introduction}
\label{introduction}
Research is a systematic, creative process that aims to generate new knowledge or apply existing knowledge in novel ways to understand, describe, explain, or predict phenomena \cite{1, 2}. It involves collecting, organizing, and analyzing data to uncover insights and support informed conclusions. Research methods can be inductive, where theories are developed based on observed patterns, or deductive, where hypotheses are tested against empirical evidence \cite{methodologies, methodologies2, methodologies3}. Research goes beyond simple discovery, as it addresses questions, solves problems, and establishes facts, contributing to the advancement of knowledge across disciplines \cite{3}.

Research methodologies and publishing practices are fundamental to advancing knowledge and fostering academic integrity \cite{4}. Since robust review and standardized evaluation criteria are crucial to ensuring reproducibility and reliability, the importance of quality assessment in research articles cannot be overstated \cite{4}. Publications serve as the primary medium through which knowledge is disseminated, having an impact on both the academic community and society \cite{publishing3, publishing, publishing2}. 

Despite significant efforts to improve methodological consistency, technical writing quality, and ethical rigor in academic publication practices, existing frameworks often lack a unified approach that addresses the diverse needs of different academic disciplines. The current literature \cite{rw1,rw2, rw3, rw4, rw6, rw7, rw9, rw10}, provides valuable guidelines but tends to focus on specific fields or aspects of the publication process, limiting their generalizability. Additionally, the ethical considerations and biases associated with Artificial Intelligence (AI)-assisted peer review and the absence of standardized templates for novice writers highlight the need for a comprehensive model \cite{6}. Inconsistencies in evaluation standards between disciplines significantly affect the credibility of research findings. As academic publishing becomes increasingly rigor-driven, there are notable gaps in achieving consistent, quality-driven manuscript preparation and evaluation processes. Despite advances in peer review, significant disparities persist in evaluation criteria between disciplines, leading to subjectivity that affects both credibility and reproducibility. This paper addresses these gaps by introducing a structured, four-dimensional framework encompassing technical content, structural coherence, writing precision, and ethical integrity. This model aims to standardize the review and publication processes in various disciplines, ensuring consistent and transparent evaluation of manuscripts, thus improving the reliability and reproducibility of research outcomes.

In this paper, general authors and reviewers are introduced to a comprehensive framework for preparing and assessing manuscripts across four dimensions, including 1) technical content, 2) structural coherence, 3) writing precision, and 4) ethical integrity. This model emphasizes the importance of novelty and validity in research, ensuring that manuscripts provide valuable insights. We then establish criteria for logical structure and coherence to facilitate reader comprehension and make the manuscript more readable. Furthermore, we recommend practices for upholding academic discourse integrity through language clarity and ethical publication. Through these objectives, the paper aspires to standardize review practices and improve the quality and societal impact of published research.

This structured approach includes the following key contributions.
\begin{itemize}
    \item Comprehensive explanation of research, research processes, methodologies, research papers, and the reviewing and publishing process.
    \item Systematic model to evaluate research innovation and rigor, ensuring meaningful contributions to the field.
    \item Criteria for organizing manuscripts to improve coherence and ease of assessment.
    \item Practical guidance for enhancing clarity and accuracy in academic communication.
    \item A framework for upholding research integrity, promoting accurate citation, and preventing unethical practices.
    \item Detailed discussion on the components of a research paper, adhering to the four-dimensional evaluation model by providing guidelines and principles.
\end{itemize}

\begin{figure*}[!h] 
    \centering 
    \begin{forest}
        for tree={
            child anchor=west,
            align=left,
            parent anchor=east,
            grow=east,
            draw,
            anchor=west,
            font=\small, 
            s sep=1mm, 
            edge path={
                \noexpand\path[\forestoption{edge}]
                (.child anchor) -| +(0.1pt,0) -- +(-2pt,0) |- 
                (!u.parent anchor)\forestoption{edge label};
            },
            edge+={thick}, 
            where level=1{tier=word}{}, 
        }
        [Holistic Approach to Research
            [5. Essential Components of a Paper\\ Adhering the Key Dimensions
                [5.14 Biography]
                [5.13 Appendix (Optional)]
                [5.12 References]
                [5.11 Acknowledgment (Optional)]
                [5.10 Conclusion and Future Work]
                [5.9 Discussion]
                [5.8 Results]
                [5.7 Methodology]
                [5.6 Related Work]
                [5.5 Introduction]
                [5.4 Keywords]
                [5.3 Abstract]
                [5.2 Authors and Affiliation]
                [5.1 Title]
            ]
            [4. Key Dimensions for Paper Preparation and Evaluation
                [4.4 Ethics in Publishing]
                [4.3 Writing Style]
                [4.2 Content Structure]
                [4.1 Technical Content]
            ]
            [3. Research
                [3.9 Editorial Board]
                [3.8 Peer Review Process]
                [3.7 Submission Process]
                [3.6 Quality of Journals]
                [3.5 Publishers]
                [3.4 Publishing Venues]
                [3.3 Papers]
                [3.2 Types]
                [3.1 Definition]
            ]
            [2. Related Work]
            [1. Introduction]
        ]
    \end{forest}
    \caption{The organizational structure of the paper.}
    \label{structure}
\end{figure*}
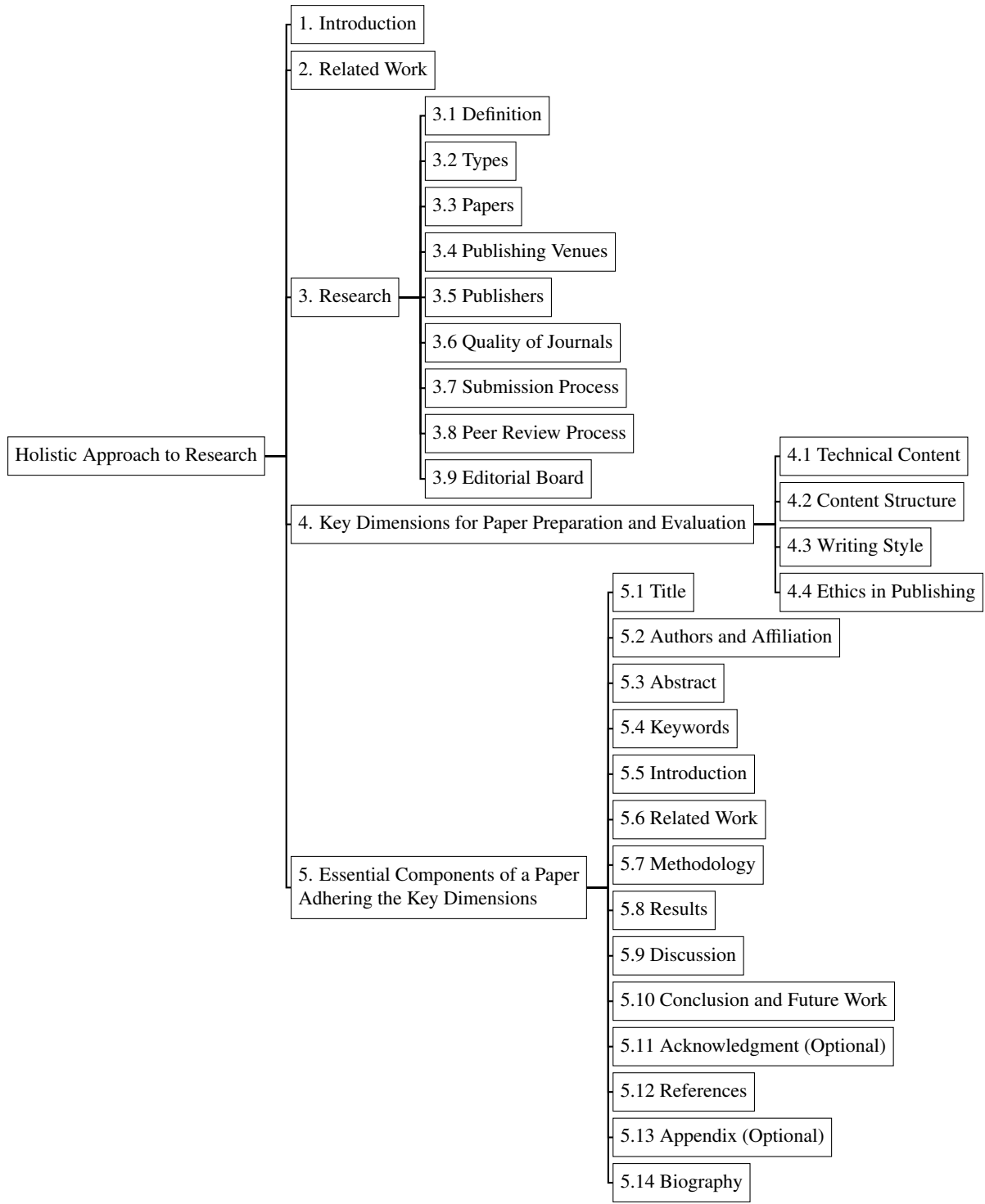

As shown in Figure \ref{structure}, the remainder of this paper is organized as follows. The literature review and related work are discussed in Section 2. A brief introduction to research, research types, research papers, and journals and publishers is presented in Section 3. Section 4 explains in detail the key dimensions for paper preparation and evaluation. Based on the key dimensions explained in the previous section, Section 5 illustrates the essential components of a research paper adhering to the key dimensions discussed in Section 4. Finally, Section 5 draws the conclusions.
\section{Related Work}

Existing research highlights significant efforts to improve methodological consistency, technical writing quality, and ethical rigor in academic publication practices. Studies underscore the importance of establishing standard frameworks that address disciplinary variability, particularly in social sciences, engineering, and biomedical research.

Aleksandrov, in \cite{rw1}, provided a comprehensive guide on the process of writing and publishing scientific manuscripts, particularly in the field of cerebrovascular diseases. In the paper, detailed instructions are provided on drafting each section of a manuscript, from the introduction and methods to the results and discussion, emphasizing clear, concise, and ethical language. The paper underscores the importance of internal review and statistical consultation in avoiding common pitfalls, such as over-interpreting results and inappropriate statistical analysis. There is, however, a potential weakness in the paper, where its focus on clinical research may limit its applicability to other scientific fields.

Feld et al., in \cite{rw2}, investigated the impact of quality writing on the perceived academic quality of economic papers. The researchers compared original and language-edited versions of PhD student papers in a randomized experiment involving 30 economists and 18 writing experts. The topics covered include the methodology of the experiment, expert evaluation criteria, and statistical analysis. The strength of the paper is its rigorous experimental design, the use of professional editors, and the demonstration that better writing significantly increases the perceived quality and acceptance chances of academic papers. However, the focus on economics papers and the specific sample of PhD students from New Zealand universities result in limited generalizability.

A comprehensive guide aimed at helping novice researchers write scientific articles was presented by Ecarnot et al. in \cite{rw3}. The paper offers practical tips to make the writing process more accessible, providing detailed instructions for structuring each manuscript section using the IMRAD format. The topics discussed are the importance of preparation, selecting a target journal, and adhering to guidelines specific to each journal. This paper emphasizes the importance of clarity and conciseness in scientific writing with its straightforward, step-by-step approach and practical advice tailored for beginners. However, the guide may be too basic for experienced researchers seeking advanced techniques and may not address challenges in highly specialized fields.

The authors in \cite{4} explored the application of AI in automating various aspects of the academic publishing process, including journal recommendation, initial quality control, reviewer selection, and post-publication quality assessment. The paper focused on the current state of AI tools in these fields, their effectiveness, and their challenges. In addition to reviewing existing AI tools and their applications and analyzing their potential benefits and limitations, the authors presented a comprehensive assessment of existing AI tools and their applications. There are, however, some weaknesses in the paper, including the lack of discussion of ethics and biases associated with AI-assisted peer review.

Rosenfeldt et al., in \cite{rw6}, provided a detailed guide for researchers, particularly in the medical field, on writing and publishing scientific papers. The paper gives practical advice on structuring a manuscript, from the introduction to the conclusion, emphasizing the importance of clear, concise writing. Besides choosing an appropriate journal, handling revisions, and dealing with rejections, it also covers practical issues. In this paper, novice writers will find help through step-by-step instructions, valuable tips, and encouragement. However, its primary focus on the medical field may limit its applicability to researchers from other fields.

Arrom et al., in \cite{rw7}, explained a detailed guide on the structure and composition of original scientific articles, emphasizing the IMRAD format (Introduction, Methods, Results, and Discussion). Practical advice is provided in each section of an article, from the title and abstract to the references, and common pitfalls were discussed. The paper offers clear, step-by-step instructions and practical tips that enhance the clarity and impact of scientific writing. A potential weakness is that it may be too general for experienced researchers looking for advanced writing techniques.

Van et al. in \cite{rw9} provided comprehensive guidance on writing literature review papers (LRPs), emphasizing the importance of adding value beyond merely summarizing existing literature. It covers the rationale for writing LRPs, different types of LRPs, methodologies for selecting papers, and how to structure the review. In addition to its detailed discussion of LRPs' added value, the paper provides practical advice on methodology and structure and includes examples in the field of transportation. The paper lacks a standardized template for novice writers, which might make it challenging for those seeking specific, field-specific advice.

The authors in \cite{rw10} presented a comprehensive guide on the principles and practices of peer-reviewing scientific manuscripts. There is a focus on the ethical considerations involved in conducting a thorough and constructive review, as well as the responsibilities of reviewers. Topics discussed include the importance of fairness, confidentiality, and integrity in the review process and practical tips for evaluating research quality, writing style, and ethical compliance. Among the most vital aspects of the paper are its detailed guidelines on reviewer conduct and its emphasis on the dual role of reviewers as advocates for authors and journals. A potential weakness of the paper is that it may not address the specific challenges faced by reviewers in highly specialized fields, which could limit its applicability to a broader audience.

Although existing research has made significant strides in improving methodological consistency, technical writing quality, and ethical rigor, several limitations remain. In many studies, particular disciplines, such as clinical research, were the focus, limiting their generalizability to other fields. Further, while some papers provide detailed guidelines on manuscript preparation and review processes, they usually lack a unified framework that addresses the diverse needs of different academic disciplines. Furthermore, ethical considerations, biases in AI-assisted peer review, and the lack of standardized templates for novice writers require further investigation and development.

To address the aforementioned challenges, this paper introduces a structured four-dimensional framework encompassing technical content, structural coherence, writing precision, and ethical integrity. This comprehensive model is designed to standardize review and publication processes across various disciplines, ensuring that manuscripts are evaluated consistently and transparently. By providing specific guidelines for each dimension, the framework helps authors prepare high-quality manuscripts that meet rigorous academic standards, thereby enhancing the reliability and reproducibility of research outcomes. Additionally, the framework promotes ethical publishing practices and reduces review bias, contributing to more trustworthy and credible academic literature.

\section{Research: Processes, Types, Papers, Reviewing, and Publishing}
\subsection{Definition}
Research is a deliberate and organized activity that is carried out to discover new knowledge, refine existing knowledge, or investigate a problem \cite{researchdef}. The research process involves formulating hypotheses, analyzing data, and drawing conclusions based on the results \cite{process}. The validity and reliability of the research are ensured by following established methods. In addition to advancing academic knowledge, research contributes to solving real-world problems in science, engineering, medicine, and social sciences.

Research begins with the identification of a problem or a knowledge gap \cite{researchgap}. This first step is crucial because it determines the direction for the rest of the research. Following this, hypotheses are developed, which provide a framework for testing predictions. A hypothesis might suggest that a new Machine Learning (ML) model would be more effective in object detection than the current models. These hypotheses lead to specific research questions that guide the study's focus and objectives.

Research requires data collection as a fundamental step. Surveys, experiments, observations, and archival research can be used to gather data depending on the field and nature of the study \cite{datacollection}. We can, for example, collect data for object detection by scraping images from the web, taking photos, and then manually or semi-automatically annotating them. Environmental scientists study the impacts of pollution on ecosystems by collecting data on air and water quality. Social scientists use surveys and interviews to gather information about human behavior and social trends.

After collecting data, patterns, correlations, and trends must be analyzed. The accuracy and reliability of this analysis are often ensured by using statistical tools and software \cite{statisticaltools}. Engineering data analysis, for instance, might involve testing materials for strength and durability under different conditions to determine their suitability for construction. Psychologists use statistical analysis to determine factors that influence mental health outcomes.

The interpretation of results is one of the most critical steps in research. In addition to comparing their findings with the original hypotheses, researchers discuss the implications of their findings. Through this interpretation, new insights can be gained and advancements can be made in the field. For example, interpreting a clinical trial's results might reveal the effectiveness of a new treatment, which medical practitioners may adopt.

Researchers follow methods and ethical standards to ensure their work's validity and reliability. The peer review process is a common practice in which other experts in the field evaluate a study's methodology and conclusions before it is published. This process ensures that the research meets high standards by identifying potential flaws. Study replication is another vital aspect, which confirms the reliability of the results by repeating the study and obtaining the same results.

Many advancements and practical applications have resulted from research in various fields. Developing new materials and energy sources are examples of technological innovations sparked by research in science. Engineers have created safer and more efficient structures and systems through research. The advancement of medical research has led to breakthroughs in treatments and vaccines that have improved public health. In addition to providing valuable insights into human behaviour and societal issues, social science research has influenced policies and practices that seek to address issues such as inequality and mental health.

Numerous examples of the impact of research can be found. COVID-19 vaccines developed quickly during the pandemic and demonstrated how critical medical research is to addressing global health crises \cite{covid}. Climate change research has been instrumental in influencing policies and practices aimed at sustainability. Research in computer science has driven advancements in artificial intelligence (AI) that have transformed industries from healthcare to finance, demonstrating the wide-reaching impacts of research \cite{ai1, ai2}. A significant milestone in machine learning (ML), a branch of artificial intelligence (AI), has been the development of deep learning (DL) algorithms, enabling computers to recognize patterns \cite{ml1}. The use of AI has enabled breakthroughs in a variety of fields, including natural language processing (NLP) \cite{nlp}, autonomous driving \cite{ad}, and healthcare diagnostics \cite{health}, demonstrating how AI can transform a variety of industries.


\subsection{Types}
There is a great deal of variation in research depending on its objective, design, and method. In this subsection, we categorize and describe different types of research according to various criteria. 

As summarized in Table \ref{research_types}, research can be classified according to various criteria, including purpose, depth of study, type of data, and methodology. The purpose of research can be categorized into theoretical and applied domains. Theoretical research focuses on generating knowledge that is not immediately applicable to practice. In contrast, applied research seeks to address specific issues by utilizing theoretical frameworks. Applied research is further classified into technological and scientific types, each aiming to enhance processes or predict outcomes.

According to the depth of the scope of the research, there are exploratory, descriptive, explanatory, and correlational research methods. Descriptive research details characteristics without probing causality, while exploratory research formulates hypotheses for under-researched topics. Correlational research examines relationships between variables, while explanation research identifies cause-and-effect relationships.

Moreover, research can be classified according to the type of data used. While qualitative research uses non-numerical data to extract meaning, quantitative research uses numerical data for statistical analysis. According to the degree of variable manipulation, research can be experimental, non-experimental, or quasi-experimental. In addition, longitudinal studies, which track changes over time, and cross-sectional studies, which assess phenomena at a specific moment, differ in timing.

The various types of research play a crucial role in advancing knowledge and addressing real-world problems. Scholars and practitioners can develop practical solutions to complex issues using multiple research methods.

\begin{table*}[!h]
    \centering
    \caption{Integrated Comparison of Research Types}
    \begin{tabular}{|p{2.5cm}|p{2cm}|p{3cm}|p{2cm}|p{1.8cm}|p{4cm}|}  
        \hline
        \textbf{Classification Metric} & \textbf{Type of Research} & \textbf{Purpose} & \textbf{Methodology} & \textbf{Data Type} & \textbf{Outcome} \\ \hline
        
        \multirow{2}{*}{\parbox{3cm}{\centering \textbf{Purpose}}} 
        & Theoretical Research & Generate new concepts & Literature review, modeling & Qualitative & Formulation of new theories and frameworks. \\ \cline{2-6}
        & Applied Research & Solving specific problems & Experimental or observational & Quantitative or Qualitative & Development of practical solutions and interventions. \\ \hline
        
        \multirow{3}{*}{\parbox{3cm}{\centering \textbf{Depth of Scope}}} 
        & Exploratory Research & Gain insights on under-researched topics & Qualitative methods & Qualitative & Identification of new research questions and hypotheses. \\ \cline{2-6}
        & Descriptive Research & Define characteristics of phenomena & Surveys, observations & Quantitative & Detailed understanding of phenomena and their characteristics. \\ \cline{2-6}
        & Explanatory Research & Establish cause-and-effect relationships & Experimental studies & Quantitative & Insights into causal mechanisms and relationships. \\ \hline
        
        \multirow{2}{*}{\parbox{3cm}{\centering \textbf{Manipulation of Variables}}}
        & Experimental Research & Test hypotheses under controlled conditions & Randomized trials & Quantitative & Validated conclusions about causal relationships. \\ \cline{2-6}
        & Non-Experimental Research & Observe phenomena in natural settings & Observational studies & Qualitative or Quantitative & Insights without manipulation, enhancing ecological validity. \\ \hline
        
        \multirow{2}{*}{\parbox{3cm}{\centering \textbf{Time Frame}}} 
        & Longitudinal Research & Monitor changes over time & Cohort studies & Quantitative & Insights into trends and long-term effects. \\ \cline{2-6}
        & Cross-Sectional Research & Observe subjects at a specific point in time & Surveys & Quantitative & Snapshot of current conditions and variables. \\ \hline
        
        \multirow{2}{*}{\parbox{3cm}{\centering \textbf{Sources of Information}}} 
        & Primary Research & Collect data directly from the source & Interviews or surveys & Qualitative & Rich, firsthand insights into specific phenomena. \\ \cline{2-6}
        & Secondary Research & Use data collected by other researchers & Literature review & Qualitative or Quantitative & Synthesis of existing knowledge and identifying gaps. \\ \hline
        
        \multirow{3}{*}{\parbox{3cm}{\centering \textbf{Inference}}} 
        & Deductive Research & Derives specific conclusions from general laws & Theoretical analysis & Qualitative & Application of theories to predict outcomes in specific cases. \\ \cline{2-6}
        & Inductive Research & Generates new theories from observations & Observational studies & Qualitative & Development of new theories based on empirical patterns. \\ \cline{2-6}
        & Hypothetical-deductive research & Tests hypotheses through deduction & Experimental studies & Quantitative & Confirmation or refutation of theoretical propositions. \\ \hline
    \end{tabular}
    \label{research_types}
\end{table*}
Research results provide insight and evidence that drive further studies and innovations, resulting in scientific advancement. Disseminating these results through journal publications, conferences, and other platforms as research papers allows the broader scientific community to review, replicate, and build upon them. By effectively disseminating research, practitioners, policymakers, and other stakeholders can benefit directly from new knowledge, helping bridge the gap between research and application.

\subsection{Research Papers}

An academic research paper communicates the results of research, experiments, or theoretical studies. Research papers advance knowledge by sharing insights and findings with the scholarly community. Research papers follow a strict structure, typically including i) Abstract, ii) Introduction, iii) Related Work, iv) Methodology, v) Results, vi) Discussion, and vii) Conclusion. 

The structured approach ensures clarity and allows readers to follow the research process and understand the findings in depth. The term `paper' refers to a document that has been accepted through peer review and published, while the term `manuscript' is used for a document that is in preparation or review.

Researchers publish research papers, both personal and professional, for several reasons. The publishing process establishes the researcher as an expert in their field and builds their academic reputation. Many academic institutions and research organizations evaluate publications when evaluating promotions, tenure, and grant applications. A researcher with a strong publication record in high-impact journals is more likely to receive funding for future projects.

In addition to fostering scientific collaborations with researchers worldwide, publishing papers can result in financially rewarding partnerships. For instance, an article on renewable energy technologies might attract the attention of industry partners looking to develop sustainable technologies.

In addition to disseminating new findings, research papers facilitate discussion on unresolved issues. They increase visibility and allow a broader audience to engage with the findings, influencing policy and practice. Public health interventions, for example, can inform government policies and improve community health.

As a result of their published work, researchers can connect with leading experts and scholars, enhancing their intellectual growth and academic network. Encouraging critical thinking and deep engagement with topics fosters continuous learning. For instance, research ideas and methodologies can be inspired by engaging with peer-reviewed literature.

The public can benefit from papers that provide insight into complex topics and contribute to evidence-based decision-making. For example, climate change research can raise awareness and lead to sustainable action. The research publication process includes communicating findings, establishing the credibility of ideas through peer review, and supporting professional growth, funding opportunities, and collaborations.


As presented in Table \ref{types_of_papers}, there are several different types of research papers, each serving a different purpose:

An \textbf{Analytical Paper} breaks a topic into parts to evaluate and draw new conclusions based on evidence. Climate change analytical papers, for instance, assess the relative effects of greenhouse gas emissions, deforestation, and industrial activities on global warming.

An \textbf{Argumentative or Persuasive Paper} presents a clear thesis and supports it with evidence to convince readers of a particular stance. For example, a persuasive paper may argue for implementing renewable energy policies using data on environmental benefits, economic feasibility, and long-term sustainability.

An \textbf{Experimental Paper} documents original experiments, describing methods, results, and analysis. This type of paper is common in the natural sciences. For example, a biology paper might explain the experimental setup, control and treatment groups, and statistical analysis of the results of a study on the effects of a new drug on cell growth.

A \textbf{Review Paper} summarizes and critically evaluates previous research to provide a comprehensive overview of a specific topic. A review paper helps identify trends, gaps, and future directions in research. For example, it is possible to synthesize findings from various studies on ML algorithms, applications, and ethical considerations in a review paper on AI.

A \textbf{Survey Paper} focuses on synthesizing trends and developments across a research field, often identifying gaps for future exploration \cite{survey}. For instance, A survey paper in computer science might examine recent advances in cybersecurity, highlighting emerging threats and innovative defences. The difference between a review paper and a survey paper is that a review paper critically evaluates and synthesizes existing research on a topic. In contrast, a survey paper summarizes and provides an overview of developments within a specific area.

A \textbf{Definition Paper} explains a concept or theory in detail, providing multiple perspectives on the topic. In philosophy, a definition paper might explore the idea of justice, examining different philosophical interpretations and their implications.

A \textbf{Compare and Contrast Paper} analyzes two or more subjects by comparing similarities and differences. This type of paper is useful for highlighting distinctions and commonalities. Literature compares and contrasts papers; for example, two novels with different cultural backgrounds might be compared based on themes and narrative techniques.
\begin{table*}[t]
\centering
\begin{tabular}{| m{3cm} | m{3cm} | m{3cm} | m{3cm} | m{3cm} |}
\hline
\textbf{Type of Paper} & \textbf{Scope} & \textbf{Objective} & \textbf{Methods} & \textbf{Average Number of Pages} \\
\hline
\textbf{Analytical Paper} & Specific topic or issue & Evaluate and draw new conclusions based on evidence & Literature review, data analysis & 10-15 \\
\hline
\textbf{Argumentative or Persuasive Paper} & Specific stance on an issue & Convince readers of a particular stance & Literature review, logical reasoning, evidence presentation & 8-12 \\
\hline
\textbf{Experimental Paper} & Original experiments & Document methods, results, and analysis & Experimental setup, data collection, statistical analysis & 12-20 \\
\hline
\textbf{Review Paper} & Comprehensive overview of a topic & Summarize and critically evaluate previous research & Literature review, synthesis of findings & 15-30 \\
\hline
\textbf{Survey Paper} & Trends and developments in a field & Identify gaps and future research directions & Literature review, trend analysis & 15-25 \\
\hline
\textbf{Definition Paper} & Concept or theory & Explain in detail, provide multiple perspectives & Literature review, theoretical analysis & 8-12 \\
\hline
\textbf{Compare and Contrast Paper} & Two or more subjects & Analyze similarities and differences & Comparative analysis, literature review & 10-15 \\
\hline
\textbf{Position Paper} & Specific issue or policy & Present an opinion or stance & Literature review, argumentation & 6-10 \\
\hline
\textbf{White Paper} & Specific problem or issue & Inform and persuade stakeholders & Case studies, data analysis & 10-20 \\
\hline
\textbf{Cause and Effect Paper} & Specific event or phenomenon & Analyze causes and effects & Literature review, data analysis & 8-12 \\
\hline
\textbf{Report} & Specific event or research findings & Present findings and recommendations & Data collection, analysis, presentation & 10-20 \\
\hline
\textbf{Interpretative Paper} & Specific text or phenomenon & Provide an interpretation or analysis & Literature review, theoretical analysis & 8-12 \\
\hline
\end{tabular}
\caption{Comparison of Different Types of Research Papers}
\label{types_of_papers}
\end{table*}

\subsection{Publishing Venues}
There are two ways in which research papers can be published: peer-reviewed or non-reviewed. In peer-reviewed publications, e.g., scholarly journals or conference proceedings, experts (peers) in the field rigorously evaluate the paper's quality, methodology, and contributions before it is accepted. Conversely, non-reviewed publications are available online through repositories, such as arXive, allowing them to be disseminated more quickly, but without the same level of scrutiny and validation as peer-reviewed publications.

The following are several possible venues where a paper can be published.

\begin{itemize}
    \item In \textbf{Scholarly journals}, peer-reviewed articles are published either in a subscription-based or open-access format. Examples include Elsevier, IEEE, Nature, Science, and The Lancet. In subscription-based journals, authors are generally exempt from article processing charges (APCs); however, some journals, such as those from IEEE, encourage authors to keep their manuscripts within a specified length, often 6–8 pages, as exceeding this limit incurs additional page fees. On the other hand, open-access journals, such as IEEE Access, make articles free to the public but require authors to cover the article processing charge. In recent years, some subscription-based journals have begun offering authors an open-access option to make their research more accessible to the public.

\item \textbf{Special Issue (SI) in Scholarly Journals}; are collections of articles focused on a specific theme or topic within a journal's scope. Guest Editors curate these issues with their expertise. An issue of a special issue highlights emerging trends, significant advances, or comprehensive reviews of a particular topic. Contributions from leading researchers provide a platform for in-depth exploration and discussion. A special issue can draw attention to cutting-edge research and foster scholarly collaboration, enhancing the journal's impact.
    
    \item \textbf{Conference Proceedings} are collections of research presented, often in volumes or as digital formats for dissemination across disciplines. As a result of these proceedings, researchers can share preliminary or ongoing research, receive peer feedback, and increase their visibility within the scientific community. Several conferences have partnerships with reputable journals, encouraging authors to submit extended versions of their accepted papers for publication. Typically, these extended submissions require additional findings, enhanced analyses, or expanded discussions to differentiate them from the original conference paper. Academic publishers such as Springer and IEEE frequently publish these proceedings, ensuring that research is accessible to a broader audience and archival in respected academic databases. In this way, conferences contribute to advancing knowledge and lay the foundation for further peer-reviewed journal publications.

    \item \textbf{Workshops} are often encouraged to be the first step for young researchers since they provide a supportive environment for early-stage research presentations and discussions. Professionals interested in specific, niche topics gather at workshops within larger conferences to share feedback and suggestions. Authors can refine their findings in a workshop based on peer feedback before submitting them to a full conference. Additional data, findings, or extended analysis are often added after a conference presentation, enabling the work to evolve and ultimately be published. With this stepwise approach, young researchers can achieve visibility, improve the rigor of their work, and increase their chances of publishing in prestigious journals.

    \item \textbf{Online repositories} like arXiv provide authors with a platform to share their research findings with the scientific community quickly. Unlike traditional journals, these repositories allow for immediate dissemination, which is especially beneficial in rapidly evolving fields. Many researchers use these platforms to gain visibility and feedback on their work, increasing the accessibility and engagement of their research. Journals permit submissions that have previously appeared in such repositories, allowing authors to share preliminary findings publicly before formal peer review. However, certain journals may view prior online availability as pre-publication and disallow submissions that have been shared in repositories. As a result of this balance, authors can disseminate their work widely while keeping options open for later publication in journals.

    \item \textbf{Books and Monographs} offer a comprehensive way to present research in-depth, making them ideal for extensive studies or multi-faceted projects. Unlike journal articles or conference papers, these are usually published as standalone works or as chapters in edited volumes. In addition to contributing to scholarly discourse, they allow researchers to present substantial analysis, theory, or methodology discussions that may not fit within shorter publication formats.

    \item \textbf{Dissertations and Theses} are required publications for obtaining a master's or doctoral degree, respectively. These works showcase a student’s research abilities, provide new insights into their field of study, and are typically subjected to rigorous review by a panel of academic advisors. Upon successful defence, these documents become part of university collections, are accessible for future reference, and may also serve as foundational material for subsequent journal articles.

    \item The goal of \textbf{institutional repositories} is to archive and disseminate research produced by university faculty, students, and researchers. By providing open access to scholarly outputs, these repositories increase the visibility and accessibility of academic work. They showcase a university's research achievements by providing free access to scholarly publications, data sets, and theses and facilitate global knowledge sharing.

    \item During a \textbf{symposium}, researchers present their findings on a particular subject or within a specialized field. Experts can discuss emerging trends and issues during these gatherings, which can be standalone events or part of larger conferences. Researchers can refine their work before submitting it to peer-reviewed venues by participating in symposia, which provide valuable networking and feedback opportunities.

    \item \textbf{Colloquium} are academic seminars held at universities where researchers, students, and faculty present their ongoing work. These events foster academic exchange, providing a forum for peer feedback and constructive discussion that can help shape future research directions. Researchers use colloquia sessions to refine their ideas and improve their presentations before presenting at conferences or publishing in journals with a broader audience.

    \item \textbf{Webinars} enable researchers to present and discuss their work with a global audience in real-time. Disseminating research to a diverse audience without travel constraints is particularly valuable in these sessions. By encouraging real-time interaction, webinars enable participants to ask questions and provide immediate feedback, making them ideal for knowledge sharing and networking.
\end{itemize}

Researchers prefer journals and conferences among the above-mentioned venues to publish their research results. It depends on several factors, such as the research stage, the desired impact, and the audience of the research paper, whether to submit it to a journal or a conference. 

A conference paper offers an opportunity to share still-in-progress research. New ideas and developments can be presented or preliminary findings can be presented. By participating in conferences, authors can receive early, informal feedback from peers, which can be helpful in refining and improving their research before submitting it to journals. Unlike journal articles, conference papers typically contain less detail and fewer references, making them more concise presentations suitable for oral or poster presentations.

On the other hand, an article in a journal is generally a fully developed summary of a researcher's work, including detailed methodology, data, and comprehensive analysis. In most cases, it presents original research results, except when it is a review paper. Data-driven papers require clear conclusions that are robustly supported by the evidence. The thorough nature of journal publications makes them highly regarded in most fields and considered significant academic achievements.

There are two types of publication options available to journals, including Open Access \cite{openaccess} and Subscription-Based Publishing. An Open Access Journal makes its content freely available to the public, ensuring a wider audience and accessibility. It is common for authors to pay an article processing charge (APC) to cover publication costs. 

Subscription-based journals restrict access to articles behind a paywall, with institutions or individuals required to purchase subscriptions. The journals generate revenue primarily through subscriptions and offer limited access without institutional or personal subscriptions.
Subscription-based models often offer rigorous editorial processes and well-established readerships, whereas open-access models promote broader reach. Depending on their goals, funding availability, and target audience, researchers must make their choice.

Academically, publishing in a journal is often considered more impressive than presenting at a conference. Typically, journals have rigorous peer-review processes and represent the top-cited articles in their fields. Citation reports also indicate that they are referenced more frequently in patents. It is, however, challenging to get published in a quality scholarly journal due to the low acceptance rates, which increases the chances of rejection.

Conversely, conferences have distinct advantages. Globally, conference proceedings are recognized as essential collections of specialized research. The conferences provide a forum for scholars to connect, exchange ideas, and collaborate. However, conferences also have limitations. Even though some conferences maintain strict selection processes, their outputs are considered less formal and permanent compared to journal publications.

Professional academic associations, universities, or commercial publishers usually publish academic journals. Within specific academic disciplines, these publishers produce peer-reviewed content that advances knowledge. A few well-known academic publishers include IEEE, Elsevier, Springer, and Wiley, which publish journals in engineering, medicine, and humanities. Many academic institutions and organizations support scholarly journals to maintain high academic rigor and promote research dissemination within particular scholarly communities.
\subsection{Publishers}
Publishers produce and distribute printed or digital content for sale or free, such as books, journals, and software, to the public. To ensure that their content is of high quality and widely accessible, publishers invest time, money, and expertise. From acquiring manuscripts and managing the editorial process to marketing and distributing the final product, this role involves a variety of tasks. Information is made available to a wide audience by publishers, who play a crucial role in spreading knowledge and culture.

In addition to facilitating knowledge dissemination, publishers provide platforms for researchers to share their findings and foster scientific discourse. Publishers ensure credibility and reliability of research by maintaining its integrity through peer review. In addition to enhancing the visibility and credibility of the authors, this process advances science and scholarship. In addition to filtering out flawed research, peer review improves the quality of published work, thus maintaining high academic standards.
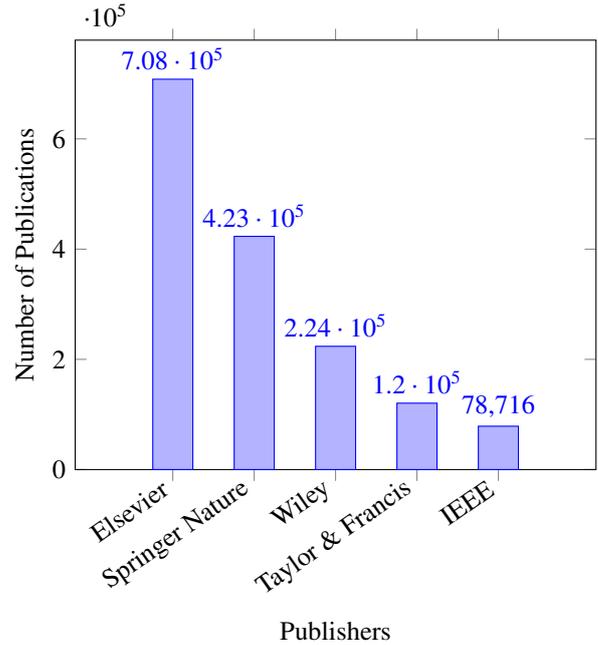
\begin{figure}[t]
    \centering
    \begin{tikzpicture}
        \begin{axis}[
            ybar,
            bar width=15pt,
            enlarge x limits=0.3,
            ymin=0,
            ylabel={Number of Publications},
            xlabel={Publishers},
            symbolic x coords={Elsevier, Springer Nature, Wiley, Taylor \& Francis, IEEE},
            xtick=data,
            xticklabel style={rotate=35, anchor=east},
            nodes near coords,
            nodes near coords align={vertical},
            every axis plot/.append style={
                fill=blue!50
            },
        ]
        \addplot coordinates {(Elsevier,708340) (Springer Nature,423210) 
                             (Wiley,223695) (Taylor \& Francis,120480) 
                             (IEEE,78716)};
        \end{axis}
    \end{tikzpicture}
    \caption{Comparison of Publication Counts Across Major Publishers As of 2024.}
\end{figure}

Some of the largest and most well-known academic publishers include:
\begin{itemize}
    \item Elsevier: Publishes journals in science, technology, and health. Elsevier is known for its extensive portfolio of high-impact journals, such as "The Lancet" and "Cell."
    \item Springer Nature: Specializes in research across multiple disciplines. It publishes influential journals like "Nature" and "Scientific Reports."
    \item Wiley: Covers fields such as science, medicine, and engineering. Wiley is recognized for its comprehensive range of academic and professional resources.
    \item IEEE: Focuses on electrical engineering and computer science. IEEE publishes leading journals and conference proceedings in these fields.
    \item Taylor \& Francis: Publishes across social sciences and humanities. It is known for its wide array of journals and books that cater to diverse academic disciplines.
\end{itemize}
In order to maximize the reach of research content, these publishers offer both subscription-based access and open access models. As open access publishing allows unrestricted access to research findings, it has gained popularity in recent years, promoting wider dissemination and impact.

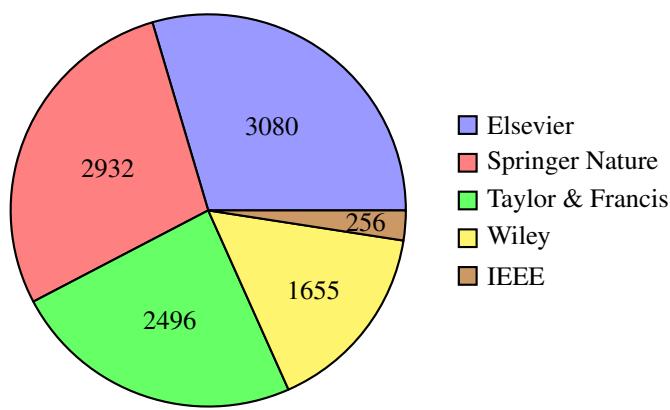
\begin{figure}[t]
    \centering
    \begin{tikzpicture}
        \pie[
            text=legend,
            radius=2.6, 
            sum=auto, 
            color={blue!40, red!50, green!60, yellow!70, brown!80} 
        ]{
            3080/Elsevier,
            2932/Springer Nature,
            2496/Taylor \& Francis,
            1655/Wiley,
            256/IEEE
        }
    \end{tikzpicture}
    \caption{Comparison of Journal Counts Across Major Publishers As of 2024.}
\end{figure}

By evaluating journals published by various publishers through its Web of Science Core Collection, Clarivate ensures they meet high standards for quality and impact. It assesses editorial rigor and best practices based on 28 criteria, categorized into quality and impact metrics. The Web of Science Master Journal List indexes and makes available journals that meet these criteria. Through Clarivate's Publisher Portal, only publishers can submit journals for evaluation.

\subsubsection{Quality of Journals}
A wide range of academic journals are published by different publishers, either focusing on highly specialized fields or taking an interdisciplinary approach. Through its Web of Science Core Collection, Clarivate Analytics categorizes journals into specific indices in order to enhance journal discovery and research impact, including:
\begin{itemize}
\item Science Citation Index Expanded (SCIE); indexes journals across various scientific disciplines, providing access to high-impact research. It enables researchers to explore leading journals in science and technology, supporting broad visibility and citation tracking for research findings.

\item Social Sciences Citation Index (SSCI); covers journals in the social sciences across 58 fields, ensuring representation of top-tier research. This index facilitates cross-disciplinary research and enhances understanding of global trends and advancements within the social sciences.

\item Arts and Humanities Citation Index (AHCI); allows researchers to locate relevant papers within arts and humanities fields with specialized precision. AHCI supports nuanced and comprehensive coverage in fields where other citation indices often fall short, emphasizing cultural and historical research relevance. \end{itemize}
These categories support subject-specific research, enabling scholars to efficiently locate journals and relevant studies across disciplines

Clarivate \footnote{\url{https://jcr.clarivate.com/}} evaluates the quality and influence of journals using a variety of metrics, including:
\begin{itemize}
\item     Impact Factor (IF); measures the average number of citations received per article published in the journal over a specific period.
The impact factor (IF) of a journal can be defined as:
\begin{equation}
    IF = \frac{{C_y}}{{N_{y-1} + N_{y-2}}},
\end{equation}
where, $C_y$ is the number of citations in current year and $N_{y-1}$ and $N_{y-2}$ are the number of published papers in the past two years.

\item Eigenfactor Score; is a reflection of the density of the network of citations around the journal using five years of cited content as cited by the Current Year. Citations are taken into account both by number and source, so highly cited sources will have a greater influence on the network than sources with fewer citations. Journal self-citations are not included in the Eigenfactor calculation.

\item Article Influence Score: Measures the average influence of a journal's articles over the first five years after publication.
\item Journal Citation Indicator (JCI): A metric introduced to provide a normalized score across fields, allowing fairer cross-disciplinary comparisons.
\end{itemize}

Journals are further grouped into quartiles (Q1, Q2, Q3, Q4) based on their performance within their subject category:
\begin{itemize}
\item Q1: Top 25\% of journals in the category.
\item Q2: 26-50\% percentile range.
\item Q3: 51-75\% percentile range.
\item Q4: Bottom 25\%.
\end{itemize}
Being published in a Q1 journal is considered prestigious, indicating high impact and scholarly recognition.

\subsubsection{Quality of Conferences}
It is essential for researchers to attend academic conferences in order to present findings, network, and keep up with advancements in their fields. Prestige and impact can be determined by the quality and rank of a conference \footnote{http://www.conferenceranks.com/}. The following factors can be used by authors to evaluate the quality of a conference:
\begin{itemize}
\item Citation Rates: High citation rates of presented papers suggest impactful research.
\item Submission Rates: High submission volumes indicate the conference's popularity and competitiveness.
\item Acceptance Rates: Lower acceptance rates imply a rigorous review process, marking the conference as selective.
\item Organizer Reputation: Conferences by reputable organizers usually draw quality submissions.
\item Research Speaker Track Records: Conferences with well-cited experts signal higher quality.
\item Ranking Databases: Platforms like Scimago and CORE rank conferences based on various metrics.
\item Participant Reviews: Feedback from attendees can provide insights into the event's organization and quality.
\item Similar Conferences: Comparing citation, acceptance rates, and organizer visibility clarifies the conference's rank in the field.
\item Institutional Support: Conferences supported by prestigious institutions are generally credible.
\item Expert Opinions: Experts can guide on conference quality based on their experience and knowledge.
\end{itemize}

Researchers can use these indicators to align their work with conferences that represent their desired visibility and rigor.

\subsubsection{Submission Process}
The choice of venue for manuscript submission is crucial. Submitting a low-quality manuscript to a high-quality journal can result in lengthy delays before rejection while submitting a high-quality manuscript to a low-quality journal diminishes the research's impact and value. Therefore, where to submit must be wisely chosen.

A manuscript should only be submitted to one journal or conference when deciding where to submit it. Multiple submissions to multiple journals are unethical and can result in rejection or blacklisting from publishers.

References cited in the manuscript may serve as a starting point for the authors to select a target journal or conference to submit a manuscript. Frequently referenced journals can serve as potential publication venues because they align with the topic. 
It can be helpful to consult supervisors or colleagues with published experience for recommendations on suitable journals or conferences.

Authors should review potential target journals or conferences' ``Aims and Scope" before submitting their articles. Manuscripts that fit within the scope of a journal or conference have a better chance of being accepted. Typically, journals and conferences announce their "Aims and Scope" on their websites.

In case of journals, several tools are available to assist the authors in identifying appropriate journals. The Elsevier Journal Finder Portal \footnote{\url{https://journalfinder.elsevier.com/}}, the Wiley Journal Finder Portal \footnote{\url{https://journalfinder.wiley.com/search?type=match}}, the Springer Journal Suggester \footnote{\url{https://journalsuggester.springer.com/}}, and IEEE Publication Recommender \footnote{\url{https://publication-recommender.ieee.org}} are effective resources that suggest relevant journals based on the manuscript's keywords or abstract.

Lastly, the authors can evaluate key metrics like journal impact indicators, review time, and conference rank in case of a conference submission. The selected journal or conference should meet the authors' academic relevance and publication timelines. These steps can optimize the submission process and increase the likelihood of publication success.

Nowadays, most journals and conferences have specific websites for manuscript submission. However, there may still be some journals without a dedicated website, requiring authors to send their manuscripts via email. Online submission systems are convenient for submitting and tracking manuscripts. For instance, IEEE uses the ScholarOne Manuscripts portal\footnote{\url{https://mc.manuscriptcentral.com/}}, while Elsevier journals primarily rely on the Editorial Manager website\footnote{\url{https://www.editorialmanager.com/}}.

Some journals only require a PDF file of the manuscript to initiate the review process. Others may request additional materials, such as a cover letter, highlights, and a graphical abstract.

The portals may ask for ORCID\footnote{\url{https://orcid.org/signin}} during the submission process. ORCID provides a persistent digital identifier (an ORCID iD) that you own and control, distinguishing you from every other researcher. You can connect your iD with your professional information, affiliations, grants, publications, peer review, and more. You can use your ID to share your information with other systems, ensuring you get recognition for all your contributions, saving you time and hassle, and reducing the risk of errors.

A Conflict of Interest (CoI) declaration is a critical statement that authors must include when submitting a manuscript to ensure transparency and integrity in scholarly publishing. CoI refers to situations where personal, financial, or professional relationships could influence, or appear to influence, the research process, analysis, or conclusions. Declaring CoI is essential to maintain the credibility of the research and trust in the publication process. Common types of CoI include financial interests (e.g., funding, grants, or payments from organizations), non-financial interests (e.g., personal relationships or affiliations), academic conflicts (e.g., competition among researchers), and institutional conflicts. Authors are required to disclose any such conflicts to allow readers and reviewers to evaluate the research objectively, ensuring that the work remains impartial and unbiased.

After submission, Elsevier journals provide a tracking system for authors to monitor their submitted manuscripts\footnote{\url{https://track.authorhub.elsevier.com/}}.

The manuscript submission process for conferences often utilizes platforms like EDAS\footnote{\url{https://www.edas.info/}}, which is widely used for managing the complete conference lifecycle. Authors submit their manuscripts through EDAS, where they are reviewed by experts in the relevant field. The platform facilitates discussions among reviewers and program committees to assess submissions and make decisions regarding acceptance or rejection. Upon acceptance, authors can complete their registration directly through EDAS, streamlining the integration of submission and registration processes. Additionally, the platform handles copyright agreements for accepted papers and supports applications for travel grants, where applicable. Authors should ensure adherence to conference-specific guidelines regarding formatting, deadlines, and other submission requirements.

\subsubsection{Peer Review Process}
Prior to publication in an academic journal, peer review is an essential step in assessing the quality of a manuscript. As a result, the research is validated, relevant, and contributes to the field according to established standards. 

There are several types of reviews, as follows. In a single-blind review, reviewers remain anonymous to authors, promoting impartial feedback. Double-blind reviews enhance anonymity by concealing both authors' and reviewers' identities, thus minimizing bias based on reputation or affiliation. On the other hand, open peer review discloses both parties' identities, fostering transparency and accountability. After the manuscript is accepted, some journals offer additional layers of scrutiny through post-publication peer review, allowing the broader academic community to critique and discuss the work, thus refining its impact and reliability over time. These approaches vary in application based on the journal's preferences and the field's standards.

\begin{figure*}[htbp]
    \centering
    \resizebox{\textwidth}{\textheight}{\includegraphics{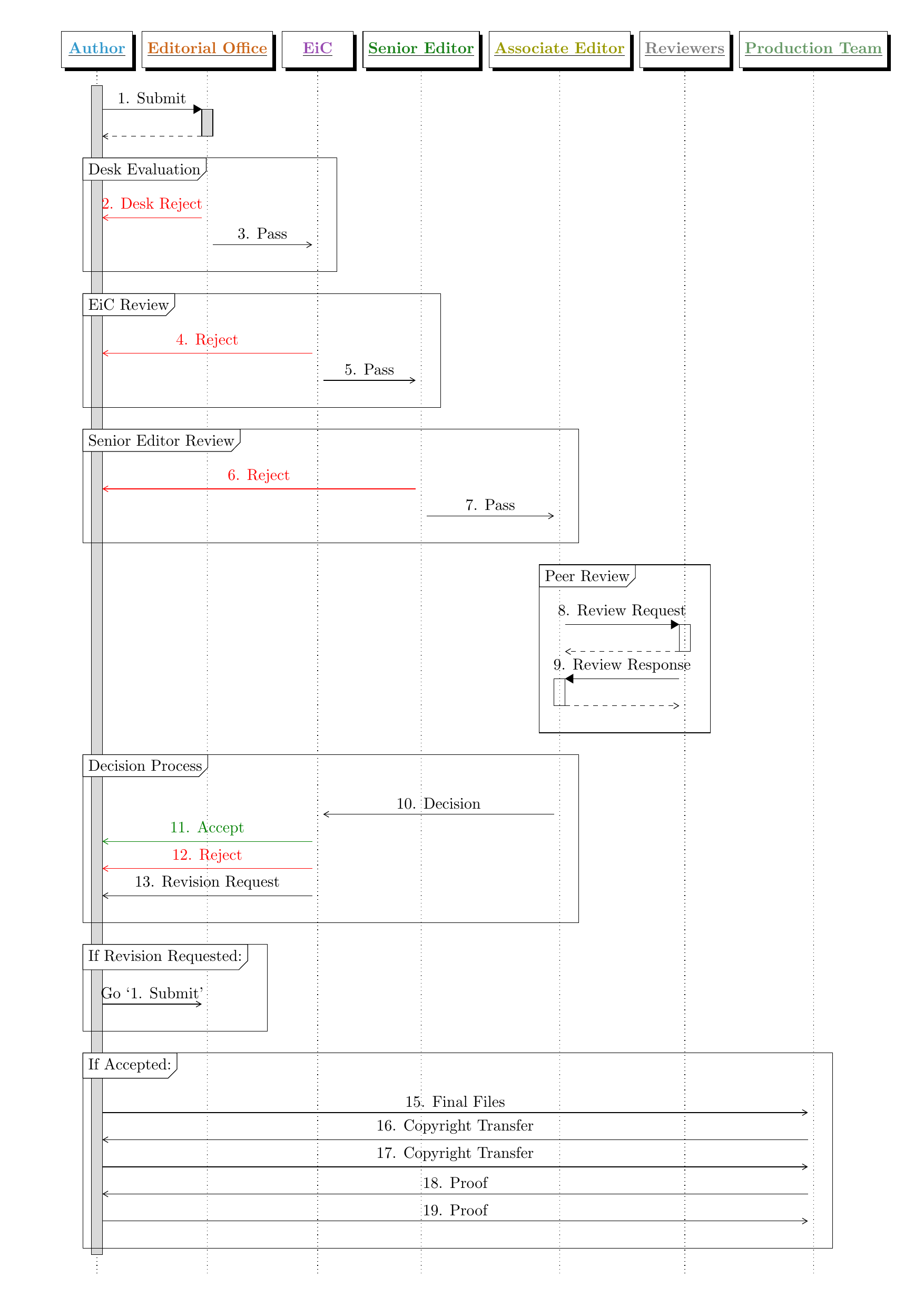}}
    \caption{Peer Review and Publishing Process of Papers}
    \label{reviewProcess}
\end{figure*}

As illustrated in Figure \ref{reviewProcess}, the stages of peer review typically include:
\begin{itemize}
 \item Submission: Researchers submit their manuscripts online through the journal's editorial office.
\item Initial Screening: The editorial office checks whether the manuscript aligns with the journal's scope and meets basic quality standards. Afterwards, the EiC passes the manuscript to a Senior Editor based on the scalability of the journal. In some journals, many submissions are received, and the EiC and Senior Editors cannot handle them all. Therefore, the Senior Editor sends the manuscript to an Associate Editor to handle the review process.
\item Reviewer Selection: The handling editor assigns the manuscript to independent experts (peer reviewers) in the relevant field.
\item Review Process: Reviewers evaluate the manuscript based on technical content, content structure, writing precision, and ethical considerations.
\item Feedback and Decision: Reviewers provide detailed feedback, e.g., a review report, recommending whether to accept, request minor or major revisions, or reject. Once the handling editor has reviewed the manuscript and received the reviewers' reports, they will make a final decision. In the case of an accepted decision, a copy of the manuscript will be sent to the Publishing department.
\item Revision and Resubmission: Authors who receive a 'Reject and Resubmit' or 'Revise and Resubmit' decision must revise and resubmit their manuscript with a response letter addressing the comments of the editor and reviewers.
The editor passes the revised revision to the reviewers after reviewing the submitted revision. Once again, the reviewers review and make a decision. Based on the recommendations of the reviewers and handling editor, the EiC makes the final decision.
\end{itemize}
By filtering out flawed work and improving quality through expert feedback, this rigorous evaluation process enhances the credibility of published research. As a result, it contributes to the integrity of academic publishing and fosters trust among scientists. This process is handled by the Editorial board of a journal, as explained above.

\subsection{Editorial Board}
Typically, well-known journals have three main boards: the editorial board, the advisory board, and the peripheral board. The editorial board includes senior editors, associate editors, managing editors, copy editors, and peer reviewers. The advisory board comprises faculty, senior, and external advisors. The peripheral board comprises a graphical designer, a platform and promotions director, and a community outreach director. Due to the importance of the editorial board, we only explain about this board.

An editorial board oversees a journal's academic integrity, quality, and direction. Their responsibilities include guiding the peer review process, providing feedback to authors, and ensuring that published content aligns with the journal's standards. Keeping the journal's reputation and publishing high-quality research depends on the editorial board.
Depending on how many submissions are received, an editorial board will have a certain number of members. Journals with fewer submissions need fewer editorial boards, e.g., an EiC and reviewers. If the journal receives too many submissions, several members, including the editorial office assistant, senior editors, associate editors, area editors, and reviews, assist the EiC.

EiC selects board members (assistants, senior editors, associate editors) in consultation with the publisher, e.g., an advisory board, to ensure a diverse and expert board. Members are chosen based on their subject expertise in the journal's field, international presence and recognition in academia, and geographic diversity to reflect a global perspective. The editorial board undergoes periodic revisions every two to three years to maintain relevance and excellence. This selection process ensures that the board remains dynamic and capable of addressing emerging trends and challenges in the field. Typically, the editorial board members are as follows.

\begin{itemize}
\item \textbf{Editorial Office Assistant}: manages communications and organizes meetings and submissions for the editorial team. The Editorial Office Assistant handles new submissions, ensuring every manuscript adheres to basic technical writing principles. For academic integrity, they perform similarity checks and plagiarism screenings and verify that the submission falls within the journal's scope. Submissions that meet such conditions will be passed to the EiC. Otherwise, they will be returned to the authors.

\item \textbf{EiC}: oversees the editorial board and makes final decisions regarding article submissions. While representing the journal in external matters, the EiC ensures the journal adheres to scientific, technical, and ethical standards. When the EiC receives a manuscript, it will undergo an initial review to ensure it is technically accurate, well-written, and aligned with the journal's focus. Manuscripts that meet these criteria are forwarded to a Senior Editor for further evaluation based on the senior editor's interests. In any case, the EIC may allow the authors to address the comments and resubmit. In any other case, the authors are not allowed to submit this article to this journal anymore, and it will be rejected.

\item \textbf{Senior Editors}: oversee the overall editorial process and make key decisions regarding article acceptance. Additionally, they coordinate the work of Associate Editors and oversee major editorial initiatives. Senior Editors evaluate manuscript quality, technical content, and writing style when they receive a manuscript. Manuscripts that do not meet the required standards are rejected and returned to the authors. Senior Editors assign manuscripts to Associate Editors based on their expertise and the manuscript's topic if they meet the standards.

\item \textbf{Associate Editors}: specialize in specific subject areas and oversee peer review of manuscripts under their responsibility. Their role is to ensure that the content aligns with the journal's scope and standards. Associate Editors invite multiple experts to review manuscripts, typically gathering three to five reviews. Following the receipt of the necessary reviews, the Associate Editor makes a decision on whether to accept, request minor revisions, request major revisions, or reject the manuscript. The decision is then forwarded to the Editor-in-Chief (EiC), who makes the final decision based on the reviews and the Associate Editor's recommendation.

\item \textbf{Guest Editors}: are invited to oversee Special Issues (SI) or thematic sections of the journal. Typically, they are experts in a particular field and are responsible for soliciting manuscripts, managing peer reviews, and making initial decisions on submissions. To ensure that the SI maintains the journal's standards and scope, Guest Editors work closely with the EiC and Senior Editors. Thanks to their contributions, fresh perspectives and cutting-edge research are crucial to the journal.

\item \textbf{Reviewers}: are experts who evaluate submitted manuscripts to ensure their quality, relevance, and academic integrity. The reviewers are normally given three to four weeks to assess a manuscript from various perspectives, such as the technical content, the structure, the style of writing, and ethical considerations. An Associate Editor receives a detailed review report from the reviewers, including feedback and recommendations. A manuscript is accepted, rejected, or revised (minor or major revision) based on the reviewers' evaluation.
\end{itemize}

After acceptance, a manuscript is sent to the publication team for review. As soon as the authors receive notification of acceptance, they must submit the final version of their manuscript. Some journals immediately publish the final files online, e.g., early access. Afterwards, the publication team will send the corresponding author a copyright transfer form. Authors will be asked to pay for additional page fees and open access papers if they are open access.
Next, the publication team formats and typesets the manuscript to align with the journal's style. A proof, including the final formatted version, is sent to the authors along with any formatting questions they may have. Either the authors will answer the questions or confirm the proof.
After the manuscript is confirmed, it is assigned a volume and page number and published online. On the designated publication date, it will be included in an issue of the journal.

The next section will discuss the key aspects of writing a high-quality manuscript for publication in top journals that are also considered by the EiC, Senior Editor, Associate Editors, and Reviewers while evaluating a manuscript.
\section{Four-Key Dimensions for Paper Preparation and Evaluation}
\begin{figure}
    \centering
    \includegraphics[width=\linewidth]{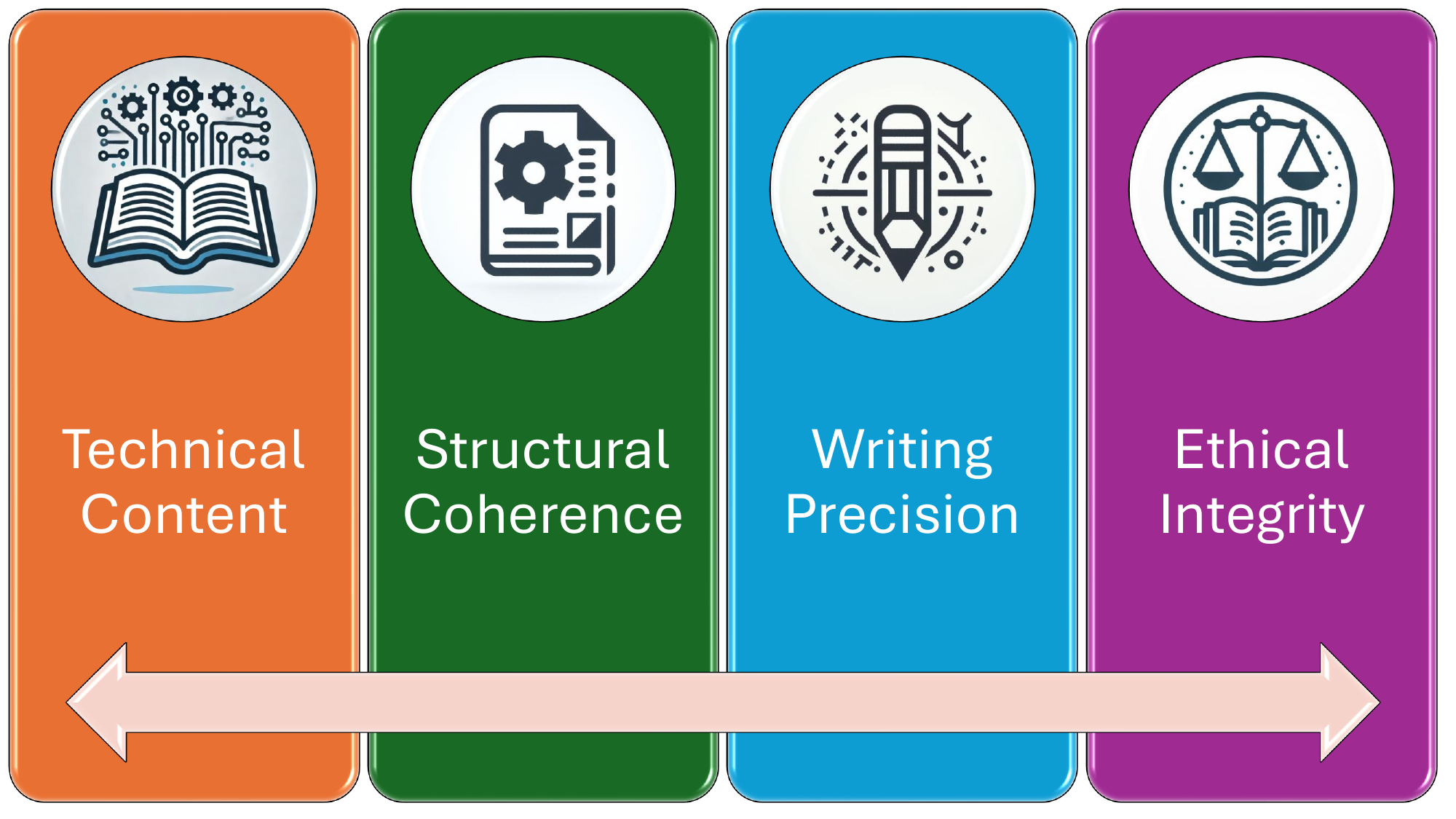}
    \caption{Four Key Dimensions for Paper Preparation and Evaluation}
    \label{dimensions}
\end{figure}

To ensure the quality and scientific integrity of a manuscript, as presented in Figure \ref{dimensions}, reviewers evaluate it according to four main dimensions, including a) technical content, b) structural coherence, c) writing precision, and d) ethical consideration. Therefore, authors should consider these four key dimensions when preparing a manuscript.

First, reviewers assess the methodology and results of the research in terms of originality, accuracy, and validity. It involves determining whether the problem statement, objectives, and technical contributions are novel and significant, whether the methods are appropriate and rigorously applied, and whether the results can be replicated and replicated. 

Second, content structure is assessed to ensure all manuscript sections are logically connected, such as the Abstract, Introduction, Methods, Results, and Discussion. A well-structured manuscript allows readers to follow the research narrative and its progression easily. 

The third dimension examines writing structure, evaluating the writing's clarity, grammar, and conciseness to ensure the manuscript is properly formatted, correctly referenced, and free of ambiguities. Research findings must be communicated in a clear and precise manner. A manuscript should be free of grammatical errors, use technical terms correctly, and adhere to the journal's style guidelines. 

Finally, there are ethical considerations to consider when writing a manuscript. Reviewers ensure proper citation practices and uphold ethical standards in academic publishing by meticulously evaluating manuscripts. Authors should not improperly use ideas, data, or materials from previous publications or from online sources without proper acknowledgment. A key component of this process is ensuring that others' words, figures, tables, and algorithms are properly referenced so that any form of plagiarism or misrepresentation can be prevented. 

Taking all of these factors into consideration ensures that the manuscript meets high academic standards and contributes meaningful insights to the field of research. To maintain the scientific record's integrity, reviewers rigorously evaluate technical content, content structure, and writing quality. In addition to improving the authors' work, this process contributes reliable and valuable knowledge to the broader academic community.

We will detail the four key dimensions of preparing and evaluating research papers in the following subsections.

\subsection{Technical Content}

The technical content of a research manuscript should clearly articulate the central scientific inquiry by including five essential elements: 1) the problem statement, 2) objectives, 3) contributions, 4) significance, and 5) merit. The problem statement answers "why" by explaining the necessity of the research. The objectives clarify "what" the research aims to accomplish. Contributions address "how" the research will be conducted, outlining methods and approaches. Significance identifies the advantages of the research compared to the related work, while merit emphasizes its value, originality, and potential advancements within the field. Together, these elements provide a comprehensive framework for understanding the research's purpose, approach, and significance.

We explain the five elements as follows.

The manuscript begins with an articulated problem statement, defining the specific issue or research gap to be addressed. It involves thoroughly reviewing existing literature to identify inadequacies or unexplored areas that require further study. It is essential to understand the research gap because it determines the relevance and contribution of the manuscript. In a study on renewable energy, for instance, it might be found that solar panels perform better under varying weather conditions than normal conditions.

The objectives should be articulated immediately after the problem statement to clarify how the research addresses identified gaps in the field. This sequence, problem statement, literature review, research gap, and objectives—create a logical flow. The literature review establishes what has been studied, identifies gaps, and provides a basis for the research need. The objectives then define the study's purpose, whether it involves developing a new framework, suggesting a method, or solving the identified problem, ensuring the study's contributions are clear and purpose-driven

Next, the technical approach to solving the problem should be clearly outlined, detailing the problem's formulation and specifying if a new algorithm or model has been developed. These technical contributions, comprising the novel core ideas of the research, should be distinct from existing work and form the foundation of the study. Additionally, the methodology should include a robust evaluation process that benchmarks the proposed solution against existing models, using empirical results, statistical analyses, or experimental validation. This approach demonstrates the model's strengths in performance metrics like accuracy, efficiency, or scalability. For instance, a new machine learning model for image recognition could be compared to existing models to highlight its improvements in accuracy and speed.

The manuscript must present comprehensive results beyond reporting outcomes to substantiate claims of superiority. Furthermore, it should explain the significance and implications of the proposed model. Results should be detailed enough to demonstrate how the proposed approach effectively addresses the research problem based on the problem statement, objective, and technical contributions. Additionally, these results should be positioned in the context of related work, demonstrating the model's advantages through relevant evaluation metrics, such as accuracy, efficiency, scalability, or robustness. The comparative analysis should highlight the model's unique strengths and demonstrate its practical value and potential for advancement.

The manuscript should then comprehensively report the merits of the research, specifically addressing the practical applications and potential impact of the findings. The purpose is to demonstrate the model or framework's usefulness and broader significance across the scientific community and potential industry applications in real-world situations. By exploring these aspects, the authors provide readers with a clear understanding of how their research could drive advancements in specific fields or lead to societal benefits. For instance, research on a novel drug delivery system might emphasize its potential to enhance patient outcomes in managing chronic diseases, offering improved therapeutic efficiency and reduced side effects. A new machine learning algorithm for predictive maintenance could also be highlighted for its potential to reduce operational costs and prevent equipment failures. Discussions of this type add depth to the manuscript by illustrating the broader value of the research beyond its theoretical contributions.

In summary, the technical content of a research paper forms the core of the evaluation process. Reviewers assess several key elements, including:

\subsubsection{Problem Statement}
A problem statement in a research paper identifies a gap or need in the current knowledge or practice. This is the heart of the research: identifying what's missing or what needs to be addressed.
 
A good research problem should be clear and concise, easily understandable, and free of ambiguity. It should be specific and focused, narrow enough to be manageable but broad enough to be significant. Furthermore, a good research problem needs to be researchable, which means that it can be investigated through the collection and analysis of data. Its significance is vital, addressing a gap that matters to the field and has practical importance or theoretical relevance. Innovation is also crucial, offering new insights, perspectives, or solutions, and it must be feasible and achievable within the researcher's time, resources, and skill constraints.

Reviewers assess the significance and relevance of a problem statement when assessing its quality. Clarity and precision are looked for, and the problem must be articulated clearly and precisely. The context and background of the problem are analyzed to establish its importance. Additionally, reviewers ensure that the problem statement leads logically to specific research questions. Literature reviews should be scrutinized to ensure the research adequately supports the problem statement. Finally, they evaluate the potential impact of solving the problem on the field or society.

Table \ref{gaps} outlines seven research gaps that can arise in academic investigations. The Theoretical Gap refers to discrepancies between observed phenomena and existing theories, often requiring further study. Methodological gaps result from insufficient or inadequate methods to answer research questions, necessitating methodological refinements. In the Empirical Gap, a lack of critical data prevents us from understanding or explaining phenomena; we can fill the data gap by conducting various studies. A Conceptual Gap arises when definitions or frameworks are not standardized due to a lack of clarity or consensus. In the Temporal Gap, research is outdated, causing discontinuities in understanding, and emphasizes the need for new studies that reflect recent advancements. A spatial gap is a lack of research in specific geographic areas, which limits generalizability and requires the inclusion of underrepresented areas in research. Last but not least, the Literature Gap reflects a failure to adequately build upon existing knowledge, which can be mitigated by thorough literature reviews and integrating recent advances into new research.  

Identifying a research gap, formulating a problem statement, and generating novel ideas to address it are essential steps in the research process. The primary challenge lies in uncovering these gaps, defining problem statements, and, more crucially, developing innovative solutions to address the identified issues.

Researchers begin by extensively examining existing literature to understand what has been done and identify gaps. This involves reading journal articles, conference papers, and reviews to see where there is a lack of data, conflicting results, or unexplored areas. Through critical literature analysis, researchers pinpoint areas that have not been fully addressed or where existing solutions are inadequate. This might involve questions like "What are the limitations of current methods?" or "What issues have not been solved by current research?"

To ensure the problem is significant and relevant to their field, researchers consider the practical applications of solving this problem. For instance, improving an existing technology's efficiency in terms of energy consumption. A clear, concise problem statement is then formulated, articulating the gap and the need for research. It should answer the "what," "why," and "how" questions succinctly.

Researchers begin by extensively reviewing existing literature to understand what has been done and identify gaps. This involves reading journal articles, conference papers, and reviews to see where there is a lack of data, conflicting results, or unexplored areas. Through critical literature analysis, researchers pinpoint areas that have not been fully addressed or where existing solutions are inadequate. This might involve questions like "What are the limitations of current methods?" or "What issues have not been solved by current research?" Researchers need to ensure that the problem is significant and relevant to their field, which often involves considering the practical applications of solving this problem. For instance, improving an existing technology's efficiency or solving a persistent health issue.
Studying survey papers is beneficial in identifying a research gap since they analyze and discuss many related topics in detail. Furthermore, survey papers often include a section on Open Research Problems for future researchers.

\begin{table*}[!h]
\begin{tabular}{|m{3cm}|m{4cm}|m{4cm}|m{5cm}|}
\hline
\textbf{Type} & \textbf{Definition} & \textbf{Examples} & \textbf{How to Address} \\
\hline
Theoretical Gap & Discrepancy between existing theories or models and observed phenomena. & Lack of studies examining the relationship between $X$ and $Y$. & Conduct further research to test existing theories or develop new theoretical frameworks. \\
\hline
Methodological Gap & Insufficiency or inadequacy in the methods used to investigate a research question. & Absence of studies utilizing qualitative methods in the field. & Review and refine research methodologies, consider alternative approaches, or combine methods to address limitations. \\
\hline
Empirical Gap & Missing data or evidence needed to fully understand or explain a phenomenon. & Limited research on the long-term effects of treatment $X$. & Gather additional data through experiments, surveys, or longitudinal studies to fill gaps in knowledge. \\
\hline
Conceptual Gap & Lack of clarity or consensus regarding key concepts or definitions in the field. & Variation in definitions of ``success" across studies. & Clarify definitions through consensus-building efforts, standardization of terms, or development of clear conceptual frameworks. \\
\hline
Temporal Gap & Lack of research over a certain period, leaving a discontinuity in understanding. & Sparse studies examining the impact of recent technological advancements. & Conduct studies to address current gaps and ensure research keeps pace with the latest developments in the field. \\
\hline
Spatial Gap & Absence of research in specific geographical areas, limiting generalizability. & Few studies exploring cultural differences in consumer behavior across regions. & Expand research efforts to include underrepresented geographical areas, ensuring a more comprehensive understanding of the phenomenon. \\
\hline
Literature Gap & Failure to address existing knowledge gaps or build upon prior research adequately. & Neglecting to consider recent advancements in the literature when designing a study. & Conduct a thorough literature review to identify existing gaps and build upon prior research findings to contribute to the advancement of knowledge. \\
\hline
\end{tabular}
\caption{Various Types of Research Gaps}
\label{gaps}
\end{table*}
\subsubsection{Objectives}
The objectives of a research paper outline the aims the study intends to achieve. They provide a clear understanding of the research's purpose and direction. Well-defined objectives guide the research design, methodology, and data collection, ensuring the study stays focused and relevant.

The qualities of a good objective are specificity, measurability, achievability, relevance, and timeliness. In addition to clearly stating what the research aims to accomplish, they should include criteria for assessing progress and success. Furthermore, they must be attainable within the scope of the study, align with the research problem, and specify a timeline for achieving them.

The best way to define objectives is to be precise and use clear, concise language to avoid ambiguity. Instead of focusing on the activities involved, it's important to emphasize the research outcomes. To refine and improve the objectives, aligning them with the research problem and hypotheses is important.

Reviewers assess the quality of objectives by checking for clarity and precision, ensuring they are articulated and unambiguous. According to the scope and resources of the study, they evaluate whether the objectives are realistic and achievable considering the research problem, questions, and methodology. Reviewers also consider the significance and impact of the objectives on the field of study and ensure that they cover all aspects necessary to address the research problem comprehensively.

Research papers succeed when their objectives are clear, guiding the researcher and reviewers through the study's goals.

\subsubsection{Technical Contributions}
\begin{figure*}
    \centering
    \includegraphics[width=0.8\linewidth]{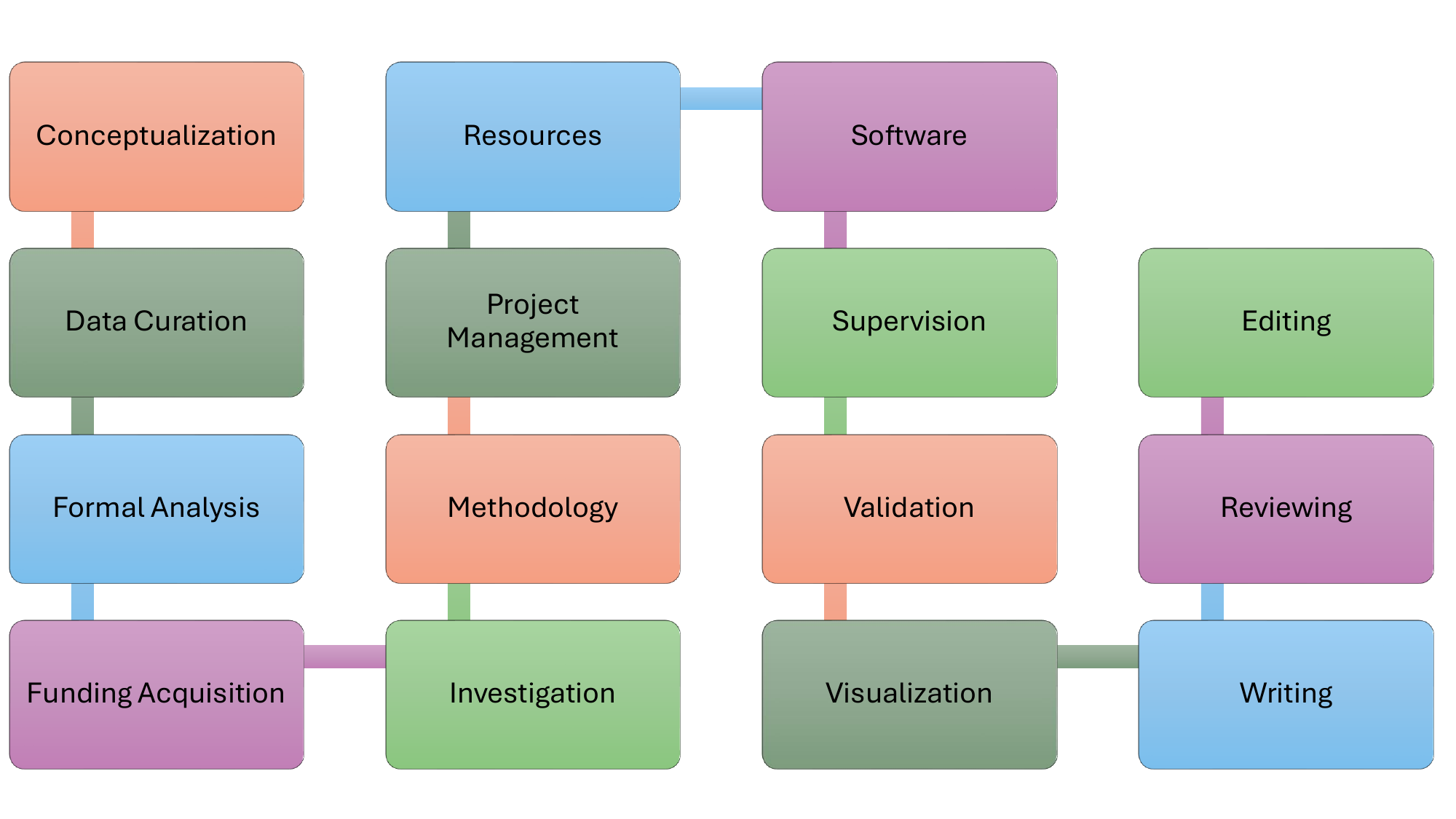}
    \caption{Possible Contributions in a Paper}
    \label{contributions}
\end{figure*}

The technical contributions of a research paper are where the author discusses their work's novel and significant contributions. Conceptualization, data curation, formal analysis, funding acquisition, investigation, methodology, project management, resources, software, supervision, validation, visualization, writing, reviewing, and editing are possible technical contributions in a paper, as presented in figure \ref{contributions}. Contributions are essential since they specify what new knowledge, methodologies, or frameworks the research offers. Originality, relevance, and impact are the defining characteristics of good contributions. They should be clearly articulated to demonstrate how the research addresses current gaps in knowledge or improves existing methods. Contributions are well-defined by detailing the unique aspects of the work and providing evidence through rigorous experiments and validation. The findings should be presented, and why they are essential and how they advance the field should be explained.

It is mandatory for each claim in the contributions to be supported by theoretical, simulation, and experimental evidence. A good research paper presents key findings, analyzes them deeply, and discusses the broader implications of these findings. To improve clarity, the results are presented structured, free of unnecessary jargon, and often supported by visual aids. Methodological analysis ensures a comprehensive understanding of the research context by uncovering patterns and relationships within the data. The discussion further integrates the findings with existing literature, explains their contribution to the field, addresses research questions or hypotheses, and discusses implications, limitations, and future directions. Together, these elements provide a coherent narrative that not only demonstrates the research's impact and relevance but also offers insights into its potential applications and contributions to advancing knowledge in the field.

Reviewers evaluate contributions according to their clarity and originality. By reviewing the research findings, they determine whether the contributions address significant gaps and problems in the field. Besides evaluating the contributions' broader impact, reviewers also consider their contribution to current knowledge or practice. Contributions that meet these criteria can convincingly demonstrate the value and relevance of the research to the academic community.

Developing new ideas is one of the most challenging aspects of research. In summary, we explain some sources for generating new ideas.
\begin{itemize} 
    \item Reading Published Papers: It is crucial for researchers to stay on top of recent publications to identify emerging trends, new methods, and areas of debate or uncertainty. A comprehensive literature review identifies research gaps and highlights potential avenues for future research. Many technical research papers define limitations and future research directions, which aid the reader in starting a new research project. Moreover, survey papers often include a section for "Open Research Problems" outlining future directions.

    \item Critical Thinking: The critical analysis of phenomena and questioning assumptions can lead to new insights and innovative solutions. Asking ``what if" and "why not" questions can spark creativity and uncover new perspectives that may not have been considered before.

    \item Attending Conferences and Workshops: 
    Conferences, workshops, seminars, webinars, and speeches are excellent opportunities for networking, hearing the latest research, and getting ideas from other peers. These events often provide inspiration and new directions for research.

    \item Discussions with Experts: Mentors, senior researchers, and industry experts can help one gain a deeper understanding of a subject matter than they can from reading alone. During these discussions, ideas can be refined and new research opportunities can be identified.

    \item Interdisciplinary Approaches: Combining methodologies and concepts from different disciplines can result in novel ideas by working with researchers from different disciplines. A cross-pollination of ideas can lead to innovative research approaches.

    \item Practical Experience: Engaging in laboratory work, conducting experiments, or engaging in fieldwork reveals practical challenges that can be addressed through research. New research questions and solutions are often inspired by real-world problems.

    \item Grant Calls and Research Proposals: Priority areas for research are highlighted by funding agencies, and reviewing these can help align research ideas with available funding. The identification of these priorities can help ensure that the research is relevant and has the potential to be funded.

    \item Teaching and Mentorship: Engaging with students and mentees can lead to fresh, thought-provoking ideas, as they often ask thoughtful questions. Additionally, teaching can provide new insights into how research can be applied in educational settings.

    \item Industry Collaborations: Using industry partnerships can provide access to data and problems unavailable in academia. Through these collaborations, direct commercial applications and research can be developed.

    \item Citizen Science and Public Engagement: Citizen science projects and public forums can reveal new research areas based on societal needs. Engaging the public can help identify community problems.
\end{itemize}

Integrating these diverse sources can generate innovative and impactful research ideas that address theoretical and practical concerns.

\subsection{Significance}
The significance of a research study refers to its importance within the existing body of work in the field. By addressing gaps in prior studies or offering a unique perspective, the paper highlights how the research contributes to knowledge advancement. Comparing the study to existing literature can help authors demonstrate how their work stands out regarding innovation, relevance, or potential impact. Clearly defining significance involves explaining why the research questions are worth exploring and how the findings may influence future research, theories, and practices. In evaluating significance, reviewers examine whether the study addresses a relevant problem, provides novel insights, improves existing knowledge frameworks or opens up new research possibilities.

\subsubsection{Merit} 
Merit emphasizes the study's direct impact and practical utility. There are many aspects to consider here, including the methodological rigor, the originality, and the potential for real-world applications. Study merit is determined by its contribution to theoretical advancements or practical solutions, making it worthwhile to academia, industry, or society. Researchers can illustrate the potential applications of their findings by describing the study's methodology, results, and implications. Study merit is evaluated by analyzing the study's methodological soundness, validity of its results, and clear demonstration of practical or theoretical benefits. Moreover, reviewers look at how well the study addresses its objectives and whether it offers useful solutions, defining it as an important contribution to the field.

A rigorous evaluation of these elements ensures that the manuscript satisfies high academic standards and contributes meaningful insights to the research community. A thorough evaluation ensures that published work is of the highest quality and maintains the integrity of scientific research.
\subsection{Content Structure}

To communicate ideas effectively and logically, a research paper must have a clear and logical structure. Conceptually, every component of a paper must be located correctly. In addition, every paragraph within each section must also be conceptually arranged correctly. A coherent narrative is created by the organization and arrangement of each section, the smooth transitions between sections of a paper, and the soft transition between paragraphs of a section. In addition to facilitating reader comprehension, well-structured papers enhance the overall impact of the research.

To maintain clarity and focus, each component of a paper, such as title, abstract, introduction, main body, and conclusion, has its own specific structure. An effective title summarizes the main theme or findings of the research, providing a clear indication of the paper's content. For example, "Federated Learning for Scalable and Real-Time Anomaly Detection in Industrial Internet of Things". The addition of "Federated Learning" clarifies the approach and innovation in the study, whereas "Industrial Internet of Things" broadens the relevance of the research by showing it applies to a significant, growing field. Finally, the terms like "Scalable" and "Real-Time" make it clear that the work addresses important performance characteristics.

An Abstract is a summary, ideally divided into three parts: the background and problem statement must be presented in the first third, while the remaining two-thirds should detail the research objectives, contributions, and significance. With this organization, readers can quickly grasp the essence of the work and are encouraged to delve deeper. 

The Introduction Section should include a contextualization of the problem, an overview of relevant literature, and a clear statement of the research question or hypothesis. For long papers, there could be a separate section after the Introduction, such as the Related Work Section. 

After that, the methodology, results, and discussion of the paper should be presented in the main body of the paper. In the main body of a paper, several sections or subsections should correspond to different areas of the research, each flowing smoothly into the next. This is crucial for maintaining the reader's interest and facilitating comprehension. For example, the methodology section should detail the problem formulation, experimental design and setup, data collection methods, and analytical techniques, ensuring the study can be replicated. The results section should present the findings clearly, often supported by tables, graphs, and statistical analyses. The discussion should interpret the results, link them to the research questions and hypotheses, and explore their implications.

Furthermore, the conclusion should encapsulate the main findings, reflect on their implications, and suggest directions for future research. In addition to reinforcing the paper's approach to the problem statement and contributing to the overall narrative, it should tie back to the introduction.

There should be a clear structure for each section of the paper, with each paragraph beginning with a topic sentence highlighting its main idea. This idea should be developed by creating logical connections to support or elaborate on the initial statement. Transitional phrases and cohesive devices ensure that sentences and paragraphs flow smoothly, much like the scenes in a well-crafted film. Transitions between sections should be smooth, with each section concluding with a sentence that connects to the next. A well-structured research paper engages readers by maintaining thematic continuity and logical progression, just as an engaging movie captivates viewers with seamless transitions and interrelated scenes.

In conclusion, a research paper is structured like a well-directed movie, with each section interconnected, each paragraph thoughtfully crafted, and every sentence contributing to the overall story. From the abstract through the conclusion, researchers can ensure their readers understand their work and its significance in the field by maintaining a clear and consistent structure.

\subsection{Writing Style}
The next key dimension for preparing and evaluating a manuscript is the writing style, which authors and reviewers must consider.
A research paper's writing style significantly impacts the clarity and effectiveness of ideas. To ensure that content is presented clearly, accurately, and easily readable, a set of guidelines must be followed. Here are key elements to consider:

The following important notes should be considered not only in the Introduction section, but throughout the manuscript.
\begin{itemize}
    \item The first time an acronym is mentioned, it must be presented in its full form.
    \item Once an acronym is defined, only the abbreviated form should be used.
    \item Well-known terminologies should be abbreviated to enhance readability.
    \item Abbreviated words should be presented in lowercase unless they are recognized terminologies (e.g., the Internet).
    \item Titles should be concise, marketable, and free of excessive clauses to enhance discoverability and engagement.
    \item Headings and subheadings should reflect the document's organization and follow proper capitalization rules.
    \item In-line lists should follow formats such as 1)... 2)... or i)... ii)...
    \item Long paragraphs should be avoided, with one idea per paragraph.
    \item Smooth transitions should be maintained between subsequent paragraphs.
    \item Numerical values less than ten should be written as words (e.g., one, two).
    \item All manuscript sections, except for "Conclusion" and "Related Work," should be written in the present tense.
    \item The manuscript should be free from writing errors, including grammatical mistakes and misplaced punctuation.
    \item A period should not be placed at the end of subsection headers.
    \item Clarity and precision should be maintained by avoiding unnecessary jargon and overly complex sentence structures.
    \item Evidence should support every claim made within the paper, with proper referencing.
    \item Figures, algorithms, and tables should be placed at the top of the page or column and referenced clearly in the text.
\end{itemize}
As a result, the writing style of a research paper plays a critical role in communicating the findings. Research can be significantly enhanced by adhering to explicit language, grammatical accuracy, structural consistency, and proper formatting. These guidelines not only assist in the presentation of the research but also align with the expectations of academic journals, thus contributing to the advancement of knowledge in the field.

\subsection{Ethics in Publishing}
The integrity and credibility of academic research depend on ethical publishing practices. Transparency, accountability, and respect for intellectual property are vital principles that authors, editors, and reviewers must follow. In the context of research, academic misconduct refers to unethical behavior that violates scholarly standards and integrity. Academic misconduct in research can take the following forms.
\begin{itemize}
    \item Plagiarism: Using someone else's research, ideas, or words without proper attribution. Taking text, data, or concepts from another source and not giving credit to the original source is considered plagiarism.

    \item Fabrication: Creating data or results that do not exist. Creating false results or inventing data can be involved in this practice.

    \item Falsification: Manipulating research data, results, or processes to fit desired outcomes. Data manipulation, misrepresentation, or selective reporting of results are examples of falsification.

    \item Authorship Misconduct: Failing to give appropriate credit to all contributors or including individuals as authors who did not contribute to the research. A ghostwriter, for example, can write a paper without getting credit for it.

    \item Duplicate Submission: Submitting the same research paper to multiple journals simultaneously without disclosing this to the publishers can lead to redundant publication and waste of peer review resources.

    \item Improper Data Handling: Not maintaining accurate and complete records of research data can prevent replication and verification of the research.

    \item Peer Review Manipulation: Attempting to influence the peer review process by suggesting favorable reviewers or suppressing critical reviews. As a result, the review process becomes less objective.
\end{itemize}

Academic misconduct in research undermines the credibility and reliability of scholarly work, damages the researchers' reputations, and erodes trust within the academic community. Institutions and publishers have strict policies and procedures to detect, investigate, and address instances of academic misconduct to uphold the integrity of scientific research.

Due to the importance of plagiarism, we will discuss it in detail in the following subsections.

\subsection{Plagiarism}
Plagiarism is a serious concern in research and other fields, such as art and technology. All academic community members need to acknowledge their debt to the authors of ideas, words, and data that form the basis of their own research.

With or without consent, plagiarism is presenting someone else's work or ideas as one's own, regardless of whether intentional, reckless, or unintentional. Materials such as photographs, songs, ideas, thoughts, expressions, and results, whether published or unpublished, may be included here. By the United States (US) Patent and Trademark Office, plagiarism is a violation of copyright laws, which protect original works of authorship, such as literary, dramatic, musical, artistic, and other intellectual works.

There are two types of plagiarism: deliberate plagiarism and unintentional plagiarism.
In deliberate plagiarism, the author copies someone else's work without giving credit. Meanwhile, unintentional plagiarism occurs when individuals are unaware of proper citation practices or don't understand plagiarism. All forms of plagiarism are considered serious offences, regardless of their intent. To avoid plagiarism, researchers must learn and adhere to proper citation and attribution rules.

Common forms of plagiarism are as follows.
\begin{itemize}
    \item Copying Without Proper Credit: Using information from the Internet or other sources without adequate referencing.
    \item Quoting Without Proper Credit: Verbatim quotations must be acknowledged with quotation marks or indentation and full source referencing.
    \item Paraphrasing Without Proper Credit: Altering a few words or the order of words from another's work without proper attribution still constitutes plagiarism.
    \item Collusion: Unauthorized collaboration between students or failure to attribute assistance received.
    \item Inaccurate Citation: Citing sources that were not actually consulted or failing to access primary sources.
    \item Patchwork Plagiarism: Combining phrases and clauses from different sources into one's own writing without proper citation.
    \item Auto-Plagiarism: Submitting identical pieces of work for multiple assessments without acknowledgment.
    \item Self-Plagiarism: Reusing portions of one's own previously published work without proper citation or acknowledgment.
\end{itemize}

Self-plagiarism might be more common among young researchers since they are unaware of the above forms of plagiarism. When an author reuses portions of their own previously published work without proper citation or acknowledgment, they may mislead readers about the novelty of the work. Common examples include duplicate publication (submitting the same or substantial parts of a manuscript to multiple journals), text recycling (reusing large sections from previous publications), data recycling (reusing previously used data without acknowledgment), and salami slicing (dividing a single substantial piece of research into multiple smaller papers). As a consequence of this practice, academic work loses its integrity, giving a false impression of the author's contributions, and academic journals and institutions generally consider it unethical. Researchers are encouraged to cite prior work appropriately and ensure each new publication contributes something unique and original to the field by citing prior work appropriately.

The consequences of plagiarism can be severe and include:
\begin{itemize}
    \item Destroyed Professional Reputation: Plagiarism can lead to losing credibility and respect within the academic community.
    \item Employment Termination: Individuals found guilty of plagiarism may be fired from their institutions.
    \item Legal Repercussions: Plagiarism can result in lawsuits, criminal charges, and monetary restitution.
    \item Loss of Publishing Privileges: Plagiarists may lose the ability to publish their work in reputable journals.
\end{itemize}

Authors should follow the following guidelines to prevent plagiarism and other unethical practices.
\begin{itemize}
    \item Properly cite and separate verbatim copied material.
    \item Obtain permission to reuse figures or tables.
    \item Accurately paraphrase and credit ideas from other sources.
    \item Familiarize themselves with publisher policies.
    \item When utilizing AI-generated content, authors must disclose its use in the acknowledgments section and cite the AI system accordingly.
    \item Authors must submit original work that has not been published elsewhere or is under simultaneous review. If a conference paper evolves into a journal article, it must ensure substantial new content is added and indicate differences from prior submissions.
    \item Submissions must not have appeared in other publications. Some journals, however, do not object if the manuscript is already available online on preprint websites like arXiv. To determine whether preprint publishing is permitted, authors should consult the target journal's website.
    \item Authors should disclose any prior rejections of their work.
    \item All previous works should be cited appropriately.
\end{itemize}

The best way to avoid plagiarism is always to give proper credit to the original authors, follow the citation style appropriate to their discipline, ensure that paraphrased content is significantly different from the original and properly cited, and get permission to use someone else's material when in doubt. Several tools, such as Turnitin \footnote{\url{https://www.turnitin.com/}} and iThenticate \footnote{\url{https://www.ithenticate.com/}}, are available to detect plagiarism and help ensure that the work submitted is original and properly cited. To maintain academic integrity and advance knowledge ethically, researchers must understand and follow these guidelines.

\subsubsection{How IEEE Detects and Deals with Plagiarism Cases?}
IEEE, one of the major publishers, defines plagiarism as the act of using another person's ideas, processes, results, or words without acknowledgment. The unethical behavior described here is a serious breach of professional ethics and cannot be tolerated. In order to categorize and evaluate the extent of plagiarism, IEEE has defined various levels of plagiarism, as illustrated in Table \ref{plagiarism_levels} \cite{ieeeplag}. The levels of plagiarism are distinguished based on the amount of material copied and how it was reproduced.

\begin{table}[t]
\centering
\begin{tabular}{|p{1.5cm}|p{6cm}|}
\hline
\textbf{Plagiarism Level} & \textbf{Description} \\
\hline
Level 1 & Copying 50–100\% of a paper is considered a flagrant violation. \\
\hline
Level 2 & 20–50\% of the content being plagiarized. \\
\hline
Level 3 & Less than 20\% of the content being plagiarized. \\
\hline
Level 4 & Improper paraphrasing, e.g., minor word changes without proper credit. \\
\hline
Level 5 & Reused material is cited but not adequately distinguished. \\
\hline
\end{tabular}
\caption{Levels of Plagiarism and Their Descriptions}
\label{plagiarism_levels}
\end{table}

When assessing possible plagiarism, evaluators consider several critical factors as follows:
\begin{itemize}
    \item Amount or Quantity of Copied Material: This includes the size of the plagiarized text, whether it is a full paper, a section, or even just a sentence.
    \item Quotation Marks Usage: Using quotation marks for direct quotes is essential to differentiate copied material.
    \item Placement of Credit Notices: Correct placement of acknowledgments ensures the original authors receive due credit.
    \item Improper Paraphrasing: Even minor changes to the original text without acknowledgment are flagged as plagiarism.
\end{itemize}

For multiple publications or submissions, IEEE requires that submitted works be original and not under consideration elsewhere. While there is no strict policy regarding how much prior work can be reused, transparency is critical. Authors must disclose any overlap and properly cite earlier works while highlighting novel contributions.

To assist in detecting plagiarism, IEEE employs the CrossCheck tool, which generates similarity reports by comparing submitted manuscripts against a vast database of published articles. These reports use color-coded matches to highlight similarities and a percentage score to indicate the extent of overlap.

The CrossCheck system \footnote{\url{https://crosscheck.ieee.org/crosscheck/}} flags manuscripts exceeding a 30\% similarity threshold for further review. When high similarity scores are identified, reviewers have three primary options:
\begin{itemize}
    \item Reject the Submission: For minor plagiarism, editors can reject the paper outright.
    \item Request Revision: Authors may be asked to revise and resubmit their manuscripts.
    \item Initiate Formal Review: Serious cases are referred to the IEEE IPR Office for a thorough investigation and appropriate action.
\end{itemize}
CrossCheck thus ensures the integrity of IEEE publications by streamlining the detection and management of plagiarism while providing necessary support for editors and volunteers.

When plagiarism is detected in a submission to IEEE, serious cases are referred to the IEEE Intellectual Property Rights (IPR) Office for a thorough investigation and appropriate action. An overview of the procedure, investigation steps, and possible actions are summarized as follows.
\begin{enumerate}
    \item \textbf{Detection:} Plagiarism is initially detected using CrossCheck, a plagiarism detection tool that compares submitted manuscripts against a large database of published technical papers and web pages.
    \item \textbf{Initial Review:} The editor responsible for the publication reviews the similarity report generated by CrossCheck.
    \item \textbf{Referral:} If the plagiarism is deemed serious, the case is referred to the IEEE IPR Office.
\end{enumerate}

There are several investigation steps as follows.
\begin{enumerate}
    \item \textbf{Appointment of Committee:} An independent committee of experts is appointed to assist in the investigation to ensure impartiality.
    \item \textbf{Review of Evidence:} The committee reviews the evidence, including the similarity report and any additional documentation provided by the editor.
    \item \textbf{Interviews:} The committee may interview the author(s) and any other relevant parties to gather more information.
    \item \textbf{Report:} The committee prepares a detailed report of their findings, including the level of plagiarism and any recommendations for corrective actions.
\end{enumerate}

If a plagiarism case is detected, several possible actions may be taken against the corresponding authors.
\begin{enumerate}
    \item \textbf{Rejection of Manuscript:} The manuscript may be rejected if plagiarism is confirmed.
    \item \textbf{Notification:} The author(s) and their institution are notified of the findings.
    \item \textbf{Author Education:} In Level 1 or Level 2 plagiarism cases, the author may be required to complete an author-education course and pass an exam.
    \item \textbf{Prohibited Authors List (PAL):} Repeat offenders may be placed on the Prohibited Authors List, which bans them from submitting new manuscripts to IEEE publications.
    \item \textbf{Notification to Vice President:} The Vice President of the IEEE Publications Services and Products Board is informed at the beginning of the investigation and after findings are reached for final approval.
\end{enumerate}

\subsection{AI for Paper Writing} 
Generative AI (Gen-AI) refers to AI models designed to produce new content based on patterns learned from large datasets, such as text, images, or code \cite{genAI}. Large Language models (LLMs) are types of Gen-AI that process and generate human language, such as GPT-3 and GPT-4 \cite{llm}. By training on a large amount of text data, LLMs can predict and generate plausible next words or sentences based on the given input (prompt).

LLMs use transformer architectures to analyze input text, understanding context and relationships between words \cite{transformers}. Based on the patterns learned during training, LLMs predict the most likely sequence of words to follow the input. It can be applied to paper writing by generating drafts, summarizing research, or improving language quality.

The use of AI models can also be integrated into systems that generate figures and equations for academic papers. However, this requires specialized tools beyond LLMs, such as those integrated with LaTeX \cite{vit}.

Furthermore, LLMs can be trained to generate references, though their reliability is a concern. Citations must be properly cited, and AI-generated references may need to be verified.

Authors may use AI at different stages as follows.
\begin{itemize}
    \item Research: LLMs can summarize academic papers, extract key information, and even suggest new avenues for research.
    \item Drafting: LLMs can help create outlines, generate ideas, and write sections of a paper.
    \item Editing: AI tools assist with grammar checks, sentence structure, and style improvements.
\end{itemize}

However, there are several concerns about using Gen-AI in paper preparation.
\begin{itemize}
    \item \textbf{Plagiarism:} 
    Unintentional plagiarism can occur, whereas AI-generated text replicates existing content. This issue arises since AI models are trained on large datasets, which may contain copyrighted material. Thus, the generated text might closely resemble or directly copy existing works without proper attribution.
    
    \item \textbf{Authorship and Integrity:} 
    Over-reliance on AI tools may blur the lines regarding a paper's true authorship. Due to this, it is challenging to determine how much of a contribution comes from humans and how much comes from artificial intelligence. The credibility of academic research depends on ensuring that the primary ideas and critical analysis are genuinely the work of the listed authors.
    
    \item \textbf{Bias:} 
    The biases found in AI models can affect the objectivity and neutrality of academic research. Biases can take many forms, including racial, cultural, or gender biases, affecting research outcomes and interpretations. These biases must be recognized and mitigated to maintain the standards of fairness and impartiality in scholarly publications.
\end{itemize}

\section{Essential Components of a Research Paper Adhering the Discussed Key Dimensions}
The previous section discussed four critical dimensions that define the quality of a research paper: technical content, content structure, writing style, and ethical concerns. This section aims to outline the essential components of a research paper with an emphasis on adherence to these criteria.

A research paper is divided into sections that describe the study's purpose, methodology, and findings. The Title summarizes the main topic, while the Authors' Affiliations indicate the researchers' affiliations. A summary of the paper's objectives, methods, and results is provided in the Abstract. Adding keywords to the document enhances discoverability by emphasizing key concepts. Following the Introduction Section, which outlines the research problem and its significance, the Related Work Section summarizes relevant literature. The next section consists of the main body, which includes problem formulation, methodology, experiment design, and setup. In the Results Section, data and findings are presented, while the Discussion interprets these findings about the research questions. The Conclusion section summarizes the research's implications and contributions. Additionally, the Acknowledgments and References sections cite all sources and express gratitude to contributors. An Appendix may also include supplementary material, such as data or methodologies, that supports the main text without being essential to its understanding. Then, a list of References is presented. In the end, some journals may ask for the photo and biography of the authors.

In the following subsections, we will explain all sections in detail, along with requirements and tips for writing each one.

\subsection{Title}
The title of a research paper encapsulates the purpose and importance of the study. An effective title should be concise yet descriptive to engage readers and convey the significance of the paper. In addition, a well-designed title increases the paper's visibility in search databases and academic repositories. Titles should indicate the unique technical contributions of the paper and distinguish it from other related papers.

\begin{itemize}
    \item The title should be concise, descriptive, and elicit curiosity, prompting the question, "Is this relevant to me?"
    \item The title is crafted to capture attention quickly, as it serves as the initial point of engagement for readers.
    \item Titles convey the study's content succinctly without using unnecessary words.    
    \item Titles should avoid uncommon acronyms unless they are widely known.    
    \item Coherence must be maintained throughout all sections of the manuscript, including title, abstract, introduction, body, and conclusion.
    \item Technical contributions are indicated in the title, e.g., main keywords, and overly broad titles are refined to focus precisely on the scope of the study.    
    \item Survey papers should have titles that accurately reflect the survey's primary contribution without being too broad.
    \item When necessary, a colon separates the title from the subtitle.    
    \item Complex sub-clauses are avoided, supporting brevity and clarity.
    \item Commonly recognized terms and abbreviations are selected to enhance readability for a broader audience.
    
\end{itemize}

\subsection{Authors' List, Affiliations, and Addresses}
To maintain transparency and increase the credibility of the research, the authors' names, affiliations, postal addresses, and e-mail addresses are included in research papers. Readers can gain a better understanding of the study's academic context and institutional backing by reading about the institutions with which the authors are associated. Through this information, readers can identify the contributing researchers and contact them for further inquiries or collaborations. 

Authors can be included in a manuscript if they have made significant contributions to its preparation. The key contributions in a research paper typically fall under the following categories: Conceptualization, Data Curation, Formal Analysis, Funding Acquisition, Investigation, Methodology, Project Administration, Resources, Software, Supervision, Validation, Visualization, Writing, Review, and Editing.

The practice of adding authors to a research paper who did not contribute is known as gift authorship or honorary authorship. When individuals are listed as authors even though they have not contributed significantly to the research, this is unethical. The reader may be misled about the actual contributors and the distribution of work within the research team.

Common reasons for gift authorship include 1) adding prominent or senior researchers to a paper to enhance its credibility or chances of acceptance (prestige), 2) including friends or colleagues to boost their academic profiles or as a return favor (favoritism), 3) listing supervisors or department heads out of obligation even if their involvement was minimal (institutional pressure), and 4) listing individuals who supported the Article Processing Charges (APC) of a paper, even though financial support alone does not justify authorship.

By distorting the actual contributions of researchers, gift authorship undermines the integrity of the research process. It is common for academic journals and institutions to have strict guidelines and criteria for authorship to prevent such practices. According to most guidelines, an author should meet all the following criteria including: 1) making substantial contributions to the conception or design of the work or the acquisition, analysis, or interpretation of data; 2) drafting or revising the manuscript critically for important intellectual content; 3) giving final approval of the version to be published; and 4) being responsible for all aspects of the work, ensuring that questions related to the accuracy or integrity of any part of the work are appropriately investigated and resolved.

Acknowledging financial support in the acknowledgments section is appropriate, but it does not qualify someone for authorship. Failure to adhere to these criteria can lead to retraction of the article, damage to the authors' reputations, and possible institutional sanctions.

For the part of authors' names, affiliations, and addresses of a paper, it is important to consider the following points.
\begin{itemize}
\item According to the contribution each author has made to the research, the order of authors is determined. Typically, the first author is the individual who has made the most significant contribution to the work.
\item Usually, the last author position is reserved for the principal investigator or senior author, who oversees the research project and mentors the team.
\item When there are multiple first authors (authors with equal contributions), a symbol is added next to their names to indicate equal contributions and a corresponding note is included to clarify their contributions (e.g., The authors contributed equally).
\item When there are multiple corresponding authors, a symbol, such as an asterisk, is placed next to their names with a footnote indicating they are the corresponding authors.
\item The affiliation information usually includes the name of the department, university, city, postal address, and country, enabling academics to network, collaborate, and recognize each other's work.
\item To maintain consistency and professionalism, authors' names, affiliations, and addresses should align with recent publications (e.g., formatting) in the target journal.
\item Authors' email addresses and ORCID \footnote{\url{https://orcid.org/}} identifiers are also included in this section to enable direct communication and to enhance the traceability of academic contributions.
\item It is essential to avoid gift authorship, where individuals who did not contribute significantly to the research are listed as authors. This includes not adding individuals who only provided financial support, such as covering the APC, as this does not qualify someone for authorship.
\end{itemize}

\subsection{Abstract}
While preparing a manuscript, its Abstract is the last part to be written. They are concise summaries that convey the essence of a research paper after other sections are completed. A particular structure is followed, which starts with a background sentence explaining what the study aims to accomplish. It is followed by a statement outlining the study's purpose. Next, the study's objectives are clearly stated, followed by a discussion of its main contributions. Finally, the abstract concludes with a statement emphasizing the significance and merit of the research. As a result of this structured approach, readers can quickly grasp the core aspects of the research without having to read the entire text.

\begin{itemize}
    \item Background, objectives, and contributions introduced in the abstract are aligned with corresponding sections throughout the paper to maintain clarity.

    \item Readers can quickly comprehend the paper's essence by understanding the research problem, methodology, key findings, and conclusions.

    \item Abstracts help readers assess a study's relevance to their interests without reading the entire paper.

    \item Abstracts are often the first thing potential reviewers and readers encounter, so clarity and precision are essential.

    \item Research findings, results, and conclusions must be summarized concisely in Abstract.

    \item Abstracts are expected to be a "stand-alone" version encapsulating the article's content and are usually between 150-300 words.

    \item Some journals require a graphical abstract for visual representation, normally a figure or diagram representing the main idea presented in the manuscript.

    \item Claims without proof are not expected in Abstracts, such as claims that the proposed model is simulated and compared with state-of-the-art methods, suggesting that all existing methods are inferior.

    \item An acceptable abstract is expected to contain one-third of the background and problem statement, with the remaining two-thirds comprising objectives, contributions, and significance.

    \item Technical contributions should be clearly stated, explaining how algorithms work and asserting their performance relative to existing methods.

    \item The ideal abstract must include background context, a problem statement, a clear objective, contributions articulated through proposals and discussions, and specific survey contributions, including problem statements and significance.

    \item It is important to articulate the advantages and significance of the proposed technique in terms of specific evaluation metrics to explain why it is superior to the related work.

    \item Justification should be given for the proposed work, along with its potential applications.

    \item Abstracts are written in the present tense to ensure that technical contributions and the significance of the proposed method are clearly stated, and comparisons are explicitly stated.
\end{itemize}

\subsection{Keywords}
Keywords are crucial to making a research paper more discoverable, as they facilitate indexing and retrieval in academic databases. Researchers usually include three to ten carefully chosen keywords summarizing the study's themes and concepts. Keywords play an important role in assigning articles to editors and categorizing them, so incorporating them into both titles and abstracts is vital. When selecting keywords, researchers should reflect on what terms they would search for and use relevant phrases instead of single words. Moreover, analyzing keywords from similar papers can provide valuable insights. The visibility and impact of a research paper can be significantly enhanced by carefully selecting and placing keywords. In summary, the following points should be considered while selecting the keyword.
\begin{itemize}
    \item Keywords facilitate efficient database indexing and enhance discoverability.
    \item The number of keywords included in a piece of writing is usually between three and ten.
    \item To improve categorization, keywords should be incorporated in the title and abstract and the rest of the manuscript.
    \item Research keywords are selected and reflected on relevant search terms.
    \item Guidance on selecting keywords can be gained from similar papers.
    \item Thoughtful keyword selection can boost visibility and citation potential.
    \item Except for proper nouns, only the first letter of keywords should be capitalized.

\end{itemize}

\subsection{Introduction}
Generally, the Introduction section of a paper introduces the paper, not the topic. Therefore, only one or two paragraphs should provide background and common knowledge about the topic, and the rest should focus on the paper.
A framework for the research is established in the Introduction Section. It begins with the background of the study, emphasizing the research problem and its significance. A review of existing literature provides context for the current study, and gaps are identified that the current study seeks to address. The literature review or related work can be a part of the introduction section or a separate section, right after the introduction section. Then, the research objectives and hypotheses are articulated, clearly stating the goals intended to be achieved through the work. Next, a clear explanation and list of contributions is presented. The structure of the manuscript is then discussed. This section is crucial for establishing the rationale behind the study.

Key Guidelines for the Introduction Section:
\begin{itemize}
    \item Introduction sections should begin with one or two paragraphs outlining the research context, closely aligned with the state-of-the-art and problem statement.
    \item A problem statement is then defined in one paragraph based on the background provided. It is the identification of an unsolved problem from the background, highlighting the main question of the research.
    \item The objective(s) are then described in a few sentences, which clarifies what the paper will investigate.
    \item Following that, contributions are explained in a few paragraphs. 
    \item The significance of the proposed method is then explained. An explanation of why this manuscript is superior to other similar published papers, such as how extensive simulations or experiments show that the proposed technique outperforms similar techniques/frameworks (in one paragraph).
    \item The paper will be easier to read with a clear list of contributions and significance. For example, "The main contributions of this paper are summarized as follows. We propose..., We present ..., We discuss..., Through extensive simulation/experiment results, we provide evidence that the proposed technique is superior to [related work] in terms of [evaluation metrics]"
    \item The merits of the proposed method should then be explained, which is a discussion of its usage and advantages in general, addressing its benefits for humanity (in one paragraph).
    \end{itemize}

\subsection{Related Work}
The "Related Work" section reviews papers within the same field that have recently been published. The related papers should be analyzed based on their objectives, contributions, results, and pros and cons. A comprehensive literature search is necessary to locate nearly all recent publications with honesty related to the topic. It may be the case that the proposed model is new, and there are no related papers on the same topic. In this case, the authors may broaden the scope and look for papers related to the topic. This section should conclude with a clear problem statement, highlighting a problem that the existing literature has not yet addressed and indicating the necessity for further investigation. In the case of a survey paper, the "Related Work" focuses solely on previously published survey papers that discuss the same topic being presented in the manuscript rather than on technical papers in the field. Including a table that contrasts the scope of the current manuscript with those of the related articles may enhance clarity for readers.

In summary, the following points should be considered in the case of the Related Work Section.
\begin{itemize}
\item The section header should read ``Related Work" or ``Literature Review".
\item "Related work" entails the study of the most recent publications in the same research domain.
\item At the end of this section, a conclusion should be drawn identifying the 'problem statement', a problem not yet considered by the related work.
\item In survey papers, "Related Work" refers exclusively to survey papers that study the same topic intended for presentation, excluding technical papers.
\item A comparison table may be included in survey papers to help readers discern the scope of this manuscript about existing related survey articles.
\end{itemize}

\subsection{Methodology}
The methodology section explains the formulation of the problem, the design and setup of the study, the data collection methods, and the analytical methods used. It is essential to provide this information for replicability and transparency so that other researchers can better understand how the study was conducted. Participants, materials, experimental procedures, and statistical analyses are typically described in this section. The validity and reliability of research findings are critically dependent on clear documentation of these methods. The name of this section should also reflect the proposed model or its abbreviation. A diagram illustrating the proposed model is also recommended to enhance understanding. There may be several subsections in this section, or it may be split into several sections.

In summary, the key components of the Methodology Section are as follows,
\begin{itemize}
    \item Problem formulation is included in this section, which identifies and clearly defines the issue the study aims to address, establishing the research's focus, scope, and relevance.
    \item A research paper on engineering may contain several mathematical equations in this section. Based on the following sentence, each equation is numbered separately and integrated into the text, concluding with either a period or a comma. Math mode should be used for in-line symbols, and bold symbols should only be used for vectors.
    \item Research design is explained as the overall strategy and framework for the study.
    \item Detailed descriptions of how data will be collected, including tools and procedures, are provided for Data Collection Methods.
    \item
    Analytical Techniques are explained by describing the statistical analysis performed on the data collected.
    \item A list of tools, instruments, or software employed in the research is presented.
    \item The experimental procedure is explained as a step-by-step account of the methodology used in the experiment.
    \item Analyses of statistics provide a clear explanation of how the data is interpreted through statistical methods.
    \item   A diagram illustrating the proposed model is also included in Diagram Representation.
    \item  The presentation of algorithms should emphasize math symbols and equations rather than long texts.
    \item   Tables must be referred to in the text according to the journal's formatting guidelines.
\end{itemize}

\subsection{Results}
Throughout the results section, the findings of the study are presented without interpretation, emphasizing their objective nature. Tables, graphs, and figures are typically included in this section to illustrate the key findings of the research. To facilitate comprehension, the data is arranged logically to enable readers to discern trends and patterns without having to delve into too much detail. The data is completely focused on what it reveals to ensure clear and effective communication of the results.

In summary, the key components of the Results Section are as follows.
\begin{itemize}
\item 
There must be some results to support every claim.
    \item Results of the study are presented without interpretation, maintaining an objective tone. Interpretation will be presented in the next section, e.g., Discussion.
    \item The data included in this report includes both quantitative and qualitative information.
    \item To illustrate key findings, tables, graphs, and figures are incorporated into the report.
    \item To enhance comprehension, the data is arranged logically in a logical manner.
    \item To make sure readers are able to understand the findings based on the findings quickly, emphasis is placed on clarity.
\end{itemize}

\subsection{Discussion}
An emphasis is placed on exploring the implications of the findings and how they relate to existing literature in the discussion section. The significance of the results, as well as potential limitations and future research avenues, are discussed in this part. The findings may be compared with previous studies, explanations for unexpected results may be proposed, and the contribution of the research to the field may be discussed. Throughout this section, the authors demonstrate critical thinking and a deep understanding of the subject matter.

Key Components of the Discussion Section are summarized as follows,
\begin{itemize}
    \item As a result of the research question, the study results are interpreted about the question itself.
    \item A broader interpretation of the findings and their implications for the future are provided.
    \item The findings are compared with existing literature to highlight similarities and differences.
    \item Several potential limitations of the study are acknowledged in the paper.
    \item Future research avenues are suggested. If there are several suggestions, there could be a separate section just before the Conclusion Section, or even the two could be mixed, e.g., Conclusion and Future Work.
    \item In the case of unexpected results, explanations are provided to explain the results.
    \item An overview of the research's contribution to the field is provided.
\end{itemize}

\subsection{Conclusion and Future Work}
This section summarizes the key findings and their implications, emphasizing the contribution of the research to the field. This section often reiterates the main points discussed in the paper and highlights the practical applications of the findings. Conclusions are crafted to be clear and concise, providing a final perspective on the research while suggesting directions for future investigations or applications. The conclusions must be consistent with the evidence and arguments presented throughout the paper. Each objective in the introduction should correspond to a concluding comment, allowing for a holistic perspective that aligns with the related work discussed. 
Additionally, authors may suggest areas for further research based on their findings in this section. A limitation of the current work can be identified; methods can be proposed to address those gaps, or topics that require further investigation can be recommended. A conclusion section highlighting future work helps readers understand where additional investigation might yield valuable insights or innovations.

Key Components of the Conclusion Section are summarized as follows.
\begin{itemize}
    \item The main findings and contributions of the research should be summarized in a clear and concise manner.
\item The significance of the research findings and their practical applications are explained.
    \item Consistently presenting evidence and arguments in the paper to support conclusions.
    \item A conclusion should be written for each objective stated in the introduction.
    \item The scope and direction of future research should be specified.
    \item Enlightenment about the importance and potential impact of the research.
    \item An overview of the paper in relation to the introduction and related work.
\end{itemize}

\subsection{Acknowledgment (Optional)}
The acknowledgment section is not mandatory and is unnecessary if the research has not received any funding from individuals or institutions. Numbering is not provided for this section, e.g., without section number. The acknowledgment section expresses gratitude to individuals and organizations that contributed to the research project. A variety of sources of support are acknowledged in this section, including funding agencies, institutions, collaborators, and individuals. It highlights the contributions of others and underscores the collaborative nature of academic research. This section usually begins with a statement of support, such as "This research was supported by...", then includes a list of acknowledgments including individuals, government agencies, grant providers and their grant numbers.

The Key Components of the Acknowledgment Section are summarized as follows.
\begin{itemize}
    \item Grants, scholarships, and institutions that provide financial assistance should be acknowledged.
    \item Those who contributed to the research but are not co-authors should be acknowledged.
    \item Resources and facilities should be acknowledged by the institution or department providing them.
    \item Supervisors, mentors, or advisors who guided the research should be appreciated, but only if they are not co-authors.
    \item Administrative staff or services that assisted in the research process should be recognized.
    \item It is optional to include thanks to friends and family who provided personal support during the research process.
\end{itemize}

\subsection{References}
There are three kinds of sentences in a research paper including: a) common knowledge, b) contributions, or 3) claims. Here is a breakdown of each as follows. Common Knowledge is widely known information that can be found across multiple sources is included here, and it can be challenging to determine who owns such words. Citations are not necessary since they are widely accepted and easily verifiable by an educated reader. For instance, a universally accepted fact or historical event falls into this category.

Claims are statements that rely on specific sources for validation, such as statistical data, findings from other research, or interpretations from specific authors or studies. To ensure academic integrity and credit the original authors, these statements must be cited.
Finally, contributions are the author's words, thoughts, insights, or conclusions that are unique to the current paper. It is important for contributions to be validated through theoretical frameworks, mathematical proofs, or empirical results presented in the study's results and discussion sections.
Classifying sources helps ensure that sources are credited appropriately and that the author's novel work is clearly identified.

The references section lists all sources cited in the research paper, formatted according to a designated citation style. This section is essential for crediting the original authors and making it easier for readers to locate the sources for further study. Including a comprehensive reference list enhances the credibility of a paper by reflecting the breadth of the research conducted. For others interested in related topics, it serves as a valuable resource. To ensure that every assertion is substantiated, each reference must provide evidence. Also, references are provided to redirect readers to read more about a topic. All cited sources must be cited in the text, with references placed at the end of sentences when they relate to the entire sentence.

Various citation styles are commonly used in academic journals and conferences, each with its own rules for formatting references. Some of the most widely used styles include:
\begin{itemize}
    \item APA (American Psychological Association): Common in the social sciences, this style emphasizes the author-date format for in-text citations and includes a detailed reference list with the author's last name, first initials, publication year, title of the work, and source. [\textit{Author, A. A. (Year). Title of the paper. Journal Name, Volume(Issue), Page numbers. https://doi.org/xxxxxx}]

    \item IEEE (Institute of Electrical and Electronics Engineers): Typically used in engineering and technology, IEEE uses numbered citations within the text, linked to a full reference list at the end. [\textit{Author First Initial., Last Name, "Title of the paper," Journal Name, vol. Volume, no. Issue, pp. Page numbers, Month Year.}]

    \item Chicago/Turabian: Two main citation methods are Notes and Bibliography (commonly used in humanities) and Author-Date (common in social sciences and physical sciences). It is flexible and used across many disciplines. [\textit{Author Last Name, First Name. "Title of the Paper." Journal Name Volume, no. Issue (Year): Page numbers. https://doi.org/xxxxxx}]

    \item MLA (Modern Language Association): Often used in the humanities, this style uses in-text citations with the author's last name and page number, accompanied by a Works Cited page. [\textit{Author Last Name, First Name. "Title of the Paper." Journal Name, vol. Volume, no. Issue, Year, pp. Page numbers.}]

    \item Vancouver: Frequently used in medical and scientific literature, this style involves numbered citations corresponding to a reference list, similar to IEEE. [\textit{Author AA. Title of the paper. Journal Name. Year; Volume(Issue) numbers. doi}]

\end{itemize}

Key Components of the References Section:
\begin{itemize}
    \item All sources must be formatted according to a specific style according to the target journal's template as discussed above (e.g., IEEE, APA, etc.).
    \item Each journal paper must include the authors' names, title, journal name, volume number, page numbers, and publication month/year.
    \item Conference references should include the authors' names, title, conference proceedings, page numbers, city, country, and year, with ``Proc." preceding the conference title.
    \item Emphasis should be placed on citing recent papers (e.g., recent three years) to reflect current research trends.
    \item Preference should be given to high-quality journal papers over online or ArXiv resources, and excessive citation of conference papers should be avoided.
    \item At least some papers published in the target journal should be cited.
    \item Titles should be placed in \{\{...\}\} to indicate acronyms in capital letters.
    \item Care should be taken to avoid citing retracted papers, ensuring the integrity of the research.
\end{itemize}

\subsection{Appendix [Optional]}
Appendices are included in research papers to provide supplementary material that is not essential to the main text but supports the findings. The purpose of this section is to provide additional details for clarity and completeness, such as lengthy data tables, complex calculations, or technical specifications that are too lengthy to be included in the main body of the paper. An appendix enhances the reader's understanding of the study without disrupting its flow. Appendices are typically labelled with letters (e.g., Appendix A, Appendix B) and referenced appropriately in the text.

Key Components of the Appendix Section:
\begin{itemize}
    \item Inclusion of additional data, tables, or figures that support the research.
    \item Provision of complex calculations or methodologies that are too lengthy for the main text.
    \item Contribution to a better understanding of the research findings without cluttering the main sections.
    \item Recognition that the appendix is not mandatory and should only be used when needed.
    \item Each appendix should be labelled alphabetically (e.g., Appendix A, Appendix B) for easy reference.
    \item Proper citation of the appendix within the main body of the paper to guide readers to supplementary information.
\end{itemize}

\subsection{Biography}
A brief author biography is typically included at the end of a research paper, providing an overview of the author's academic background, professional experience, and areas of expertise. This section highlights the author's contributions to the field, current research interests, and affiliations to establish credibility for the work presented. Readers could connect the research with the author's professional profile and expertise.

An effective author biography often includes the following elements:
\begin{itemize}
    \item Author's full name, highest academic degree, and current professional title.
    \item Name of the institution or organization where the author is employed, along with the department or division.
    \item Summary of academic history, particularly focusing on the highest degree(s) earned and the institutions attended.
    \item Brief mention of primary research interests and fields of specialization.
    \item Reference to significant contributions in the field, including high-impact publications or groundbreaking projects.
    \item Information on memberships in professional organizations, editorial roles, and any prestigious awards or honors received.
\end{itemize}

\section{Conclusion}
In this study, we introduced a structured, four-dimensional framework for preparing and evaluating research quality across diverse academic disciplines. By focusing on technical content, structural coherence, writing precision, and ethical integrity, the proposed model addressed critical gaps in current peer review practices, promoting a more transparent and consistent approach to research evaluation. In each dimension, specific guidelines were provided to improve manuscript rigor, ethical standards, and clarity, enhancing the reliability of academic publications and their societal impact. Moreover, we provided a detailed discussion on the components of a research paper, adhering to the four-dimensional evaluation model by providing guidelines and principles.
As a result of this framework, significant practical implications emerge. For authors, it provides a comprehensive guide to structuring research that meets rigorous standards, aiding in the production of high-quality, impactful publications. For reviewers, it offers a clear set of criteria to ensure consistency in assessments, fostering equitable and objective evaluations across disciplines. 

Research in the future should focus on implementing this framework across various publishing platforms and refining its components based on academic feedback. Furthermore, further studies could explore how to adapt this model to evolving ethical considerations, such as AI-generated content and open access. With these initiatives, the academic publishing field can work towards a more reliable, standardized, and ethically robust scholarly communication landscape.

\section*{Acknowledgment}
This research was supported by a grant (2022-MOIS63-001(RS-2022-ND641011)) of Cooperative Research Method and Safety Management Technology in National Disaster funded by Ministry of Interior and Safety (MOIS, Korea).

 \bibliographystyle{elsarticle-num-names}
\bibliography{08_References}





\vspace{28pt}
\begin{wrapfigure}{l}{0.17\textwidth}
    \includegraphics[width=0.2\textwidth]{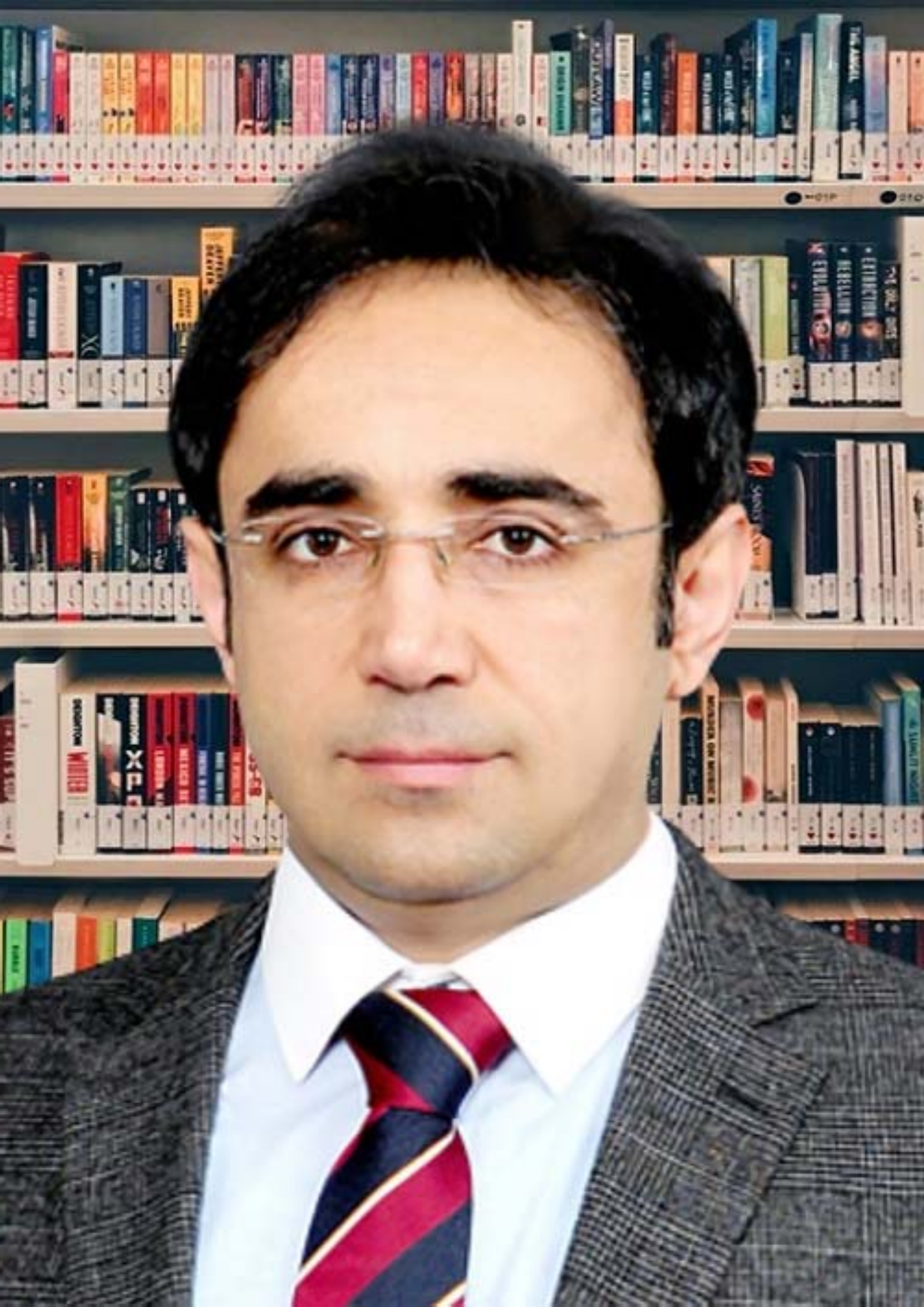}
\end{wrapfigure}
\noindent \textbf{MD. JALIL PIRAN (Senior Member, IEEE)} holds a distinguished academic background and currently serves as an Associate Professor at the Department of Computer Science and Engineering in Sejong University, Seoul, South Korea. He earned his Ph.D. degree in Electronics and Information Engineering from Kyung Hee University, South Korea, in 2016, and subsequently worked as a Post-Doctoral Fellow at the Networking Laboratory of the same institution.

Prof. Piran has made significant contributions to the field of Artificial Intelligence and Data Science through his extensive research publications in esteemed international journals and conferences. His areas of expertise encompass Machine Learning, Data Science, Big Data, Internet of Things (IoT), and Cyber Security.
In addition to his research endeavors, Prof. Jalil Piran actively engages with scholarly journals as an Editor, including the "IEEE Transactions on Intelligent Transportation Systems," "Elsevier Journal of Engineering Applications of Artificial Intelligence," "Elsevier Journal of Physical Communication," and "Elsevier Journal of Computer Communication", as well as the Guest Editor for " IEEE Transactions on Consumer Electronics". He also serves as Secretary of the IEEE Consumer Technology Society on Machine Learning, Deep Learning, and AI.Furthermore, he assumes the role of Track Chair for Machine Learning, Deep Learning, and AI in the CE (MDA) Track for the 2024 IEEE International Conference on Consumer Electronics (ICCE).In 2022, he chaired the "5G and Beyond Communications" Session at the prestigious IEEE International Conference on Communications (ICC). 
Prof. Piran is esteemed as a Senior Member of IEEE and represents South Korea as an Active Delegate to the Moving Picture Experts Group (MPEG). His outstanding research contributions have been recognized internationally, as evidenced by the prestigious "Scientist Medal of the Year 2017" awarded by IAAM in Stockholm, Sweden. Moreover, he received accolades from the Iranian Ministry of Science, Technology, and Research as an "Outstanding Emerging Researcher" in 2017. His exceptional Ph.D. dissertation was honored as the "Dissertation of the Year 2016" by the Iranian Academic Center for Education, Culture, and Research in the Engineering Group.\\\\\\

\begin{wrapfigure}{l}{0.17\textwidth}
    \includegraphics[width=0.2\textwidth]{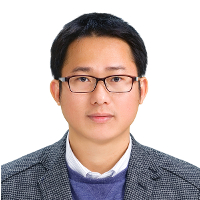}
\end{wrapfigure}
\noindent \textbf{Nguyen H. Tran} (Senior Member, IEEE) received the B.S. degree in electrical and computer engineering from the Hochiminh City University of Technology, in 2005, and the Ph.D. degree in electrical and computer engineering from Kyung Hee University, in 2011. He was an Assistant Professor with the Department of Computer Science and Engineering, Kyung Hee University, from 2012 to 2017. Since 2018, he has been with the School of Computer Science, The University of Sydney, where he is currently a Senior Lecturer. His research interest includes distributed computing and learning over networks. He received the Best KHU Thesis Award in engineering, in 2011, and several best paper awards, including the IEEE ICC 2016, APNOMS 2016, and IEEE ICCS 2016. He receives the Korea NRF Funding for Basic Science and Research, from 2016 to 2023. He has been an Editor of the IEEE Transactions on Green Communications and Networking, since 2016.
\newpage
\clearpage
\onecolumn
\section*{Appendix A: Checklists for Different Sections of a Paper adhering the Discussed Key Dimensions}
\begin{longtable}{|p{0.25\textwidth}|p{0.6\textwidth}|p{0.081\textwidth}|}
    \hline
    \textbf{Section} & \textbf{Criteria} & \textbf{\checkmark} \\
    \hline
    \endfirsthead

    \hline
    \textbf{Section} & \textbf{Criteria} & \textbf{\checkmark} \\
    \hline
    \endhead

    \hline
    \endfoot

    \hline
    \endlastfoot

    \multirow{12}{*}{Title} & Is the Title concise, descriptive, and engaging? & \\
    \cline{2-3}
    &Does the Title prompt readers to consider its relevance? & \\
    \cline{2-3}
     & Does the Title succinctly convey the study's content without unnecessary words? & \\
    \cline{2-3}
     & Are uncommon acronyms avoided, or are only widely recognized acronyms used in the Title? & \\
    \cline{2-3}
     & Is the Title consistent with the content across all sections (abstract, introduction, main body, conclusion)? & \\
    \cline{2-3}
     & Are the topics, objectives, and contributions introduced in the abstract consistent with the Title? & \\
    \cline{2-3}
     & For technical papers, does the Title indicate specific contributions and avoid broad scope? & \\
    \cline{2-3}
     & For survey papers, is the Title focused on the survey’s primary contributions, avoiding excessive breadth? & \\
    \cline{2-3}
     & Does the Title include a colon to divide the main title and subtitle for clarity, if necessary? & \\
    \cline{2-3}
     & Are complex sub-clauses avoided, supporting brevity and clarity? & \\
    \cline{2-3}
     & Are commonly recognized terms and abbreviations used to improve readability? & \\
    \cline{2-3}
     & Does the title capture attention quickly as the initial point of engagement for readers? & \\
    \cline{2-3}
     & Does the title reflect and distinguish the unique technical contributions from similar studies? & \\
    \hline \hline

    \multirow{7}{*}{Authors List and Information} & Is the order of authors determined by their contributions, with the most significant contributor listed first? & \\
    \cline{2-3}
     & Is the last author position reserved for the principal investigator or senior author? & \\
    \cline{2-3}
     & Is a symbol added for authors with equal contributions, and is a note provided to clarify this? & \\
    \cline{2-3}
     & Are corresponding authors designated with an asterisk and a footnote? & \\
    \cline{2-3}
     & Does the affiliation include the institution name, department, city, postal code, country, and email address? & \\
    \cline{2-3}
     & Is the presentation of authors’ names and affiliations consistent with recent publications in the target journal? & \\
    \hline \hline

    \multirow{15}{*}{Abstract} & Is the abstract a concise summary within 150–250 words? & \\
    \cline{2-3}
     & Does it provide a brief background or context for the study? & \\
    \cline{2-3}
     & Is the research problem clearly outlined in a single sentence? & \\
    \cline{2-3}
     & Are the main objectives of the study stated clearly? & \\
    \cline{2-3}
     & Does it summarize the methodology used? & \\
    \cline{2-3}
     & Are key findings or results of the study highlighted? & \\
    \cline{2-3}
     & Is the significance and relevance of the research clearly conveyed? & \\
    \cline{2-3}
     & Does the abstract provide a concise summary of contributions and conclusions? & \\
    \cline{2-3}
     & Is it written in the present tense? & \\
    \cline{2-3}
     & Does it stand alone without requiring additional context from the paper? & \\
    \cline{2-3}
     & Are technical contributions and their relevance to state-of-the-art methods specified? & \\
    \cline{2-3}
     & Are relevant keywords or index terms included? & \\
    \cline{2-3}
     & Is the tone of the abstract clear and precise, avoiding excessive details? & \\
    \cline{2-3}
     & Does it include potential impacts or benefits of the proposed work? & \\
    \cline{2-3}
     & Is the abstract free from overly assertive or unexpected claims? & \\
    \hline \hline

    \multirow{8}{*}{Keywords} & Are there between three to ten keywords included? & \\
    \cline{2-3}
     & Are the keywords thoughtfully selected to represent the core themes and concepts of the study? & \\
    \cline{2-3}
     & Are keywords incorporated in both the title and abstract to enhance categorization and discoverability? & \\
    \cline{2-3}
     & Are potential search terms relevant to the research reflected upon when keywords are selected? & \\
    \cline{2-3}
     & Are specific and relevant phrases chosen instead of single, general words? & \\
    \cline{2-3}
     & Does each keyword use lowercase except for the first letter of proper nouns? & \\
    \cline{2-3}
     & Is the selection of keywords likely to boost the visibility and citation potential of the paper? & \\
    \hline \hline

    \multirow{14}{*}{Introduction} & Does the introduction provide a brief background (1-2 paragraphs) outlining the research context, closely aligned with state-of-the-art and problem statement? & \\
    \cline{2-3}
     & Is there a clear identification of an unaddressed problem (problem statement) in the background section? & \\
    \cline{2-3}
     & Is the main research objective clearly stated in a single sentence? & \\
    \cline{2-3}
     & Are the key contributions explicitly listed, with phrases like ``we propose," ``we present," and ``we discuss"? & \\
    \cline{2-3}
     & Is there an explanation of the manuscript's significance, comparing its value to other similar research? & \\
    \cline{2-3}
     & Does the introduction highlight the general benefits and applications of the proposed work (1-2 paragraphs)? & \\
    \cline{2-3}
     & Are acronyms defined at their first mention, with only the abbreviated form used afterwards? & \\
    \cline{2-3}
     & Are well-known terminologies consistently abbreviated to enhance readability? & \\
    \cline{2-3}
     & Are headings and subheadings accurately reflecting the paper's organization? & \\
    \cline{2-3}
     & Does the introduction avoid long paragraphs, keeping one main idea per paragraph? & \\
    \cline{2-3}
     & Is a smooth transition maintained between paragraphs? & \\
    \cline{2-3}
     & Are numerical values less than 10 written in words? & \\
    \cline{2-3}
     & Is the introduction free from writing errors, including grammar and punctuation issues? & \\
    \cline{2-3}
     & Is the introduction written in the present tense, except when referring to prior research? & \\
    \hline \hline

    \multirow{6}{*}{Related Work} & Is the section titled ``Related Work" or ``Literature Review"? & \\
    \cline{2-3}
    & Does the section include a review of the most recent publications in the same research domain? & \\
    \cline{2-3}
     & Does it conclude with a problem statement that highlights an unresolved issue in existing literature? & \\
    \cline{2-3}
     & For survey papers, does the ``Related Work" section focus exclusively on prior survey articles and exclude technical papers? & \\
    \cline{2-3}
     & Is a comparison table included (in survey papers) to show the scope of this paper relative to other related survey articles? & \\
    \cline{2-3}
    & Is the language clear and concise, allowing readers to follow the reviewed works and understand the presented gap easily? & \\
    \hline \hline
    
    \multirow{1}{*}{Methodology} &      Is the problem clearly identified and defined?& \\
    \cline{2-3}
    & Does the problem formulation establish the research’s focus, scope, and relevance?& \\
    \cline{2-3}
    & Is the overall strategy and framework for the study explained?& \\
    \cline{2-3}
    & Is the design of the study (e.g., prospective, retrospective, randomized, controlled) clearly stated?& \\
    \cline{2-3}
    & Are any unusual methodologies justified with appropriate references or context?& \\
    \cline{2-3}
    & Are detailed descriptions of how data will be collected provided?& \\
    \cline{2-3}
    & Are the tools and procedures for data collection clearly described?& \\
    \cline{2-3}
    & Is the study population (e.g., participants, animals, cells) adequately described?& \\
    \cline{2-3}
    & Are inclusion and exclusion criteria for participants clearly stated?& \\
    \cline{2-3}
    & Are the statistical analyses performed on the collected data clearly explained?& \\
    \cline{2-3}
    & Is the primary endpoint specified, along with the methods used to measure it?& \\
    \cline{2-3}
    & Are secondary endpoints and their measurement methods described?& \\
    \cline{2-3}
    & Is a list of tools, instruments, or software employed in the research provided?& \\
    \cline{2-3}
    & Are manufacturer details (name, city, country) for specific equipment or tests included?& \\
    \cline{2-3}
    & Is the experimental procedure explained as a step-by-step account?& \\
    \cline{2-3}
    & Are all interventions, operations, questionnaires, and imaging techniques detailed?& \\
    \cline{2-3}
    & Are ethical considerations, such as ethics committee approval and informed consent, addressed?& \\
    \cline{2-3}
    & Are standard statements about data presentation included (e.g., means, standard deviation, median)?& \\
    \cline{2-3}
    & Are specific statistical approaches listed for each type of variable?& \\
    \cline{2-3}
    & Is the sample size justification provided, including working hypothesis, expected differences, and alpha/beta risks?& \\
    \cline{2-3}
    & Are planned subgroup analyses detailed to avoid criticisms about post hoc studies?& \\
    \cline{2-3}
    & Is a diagram illustrating the proposed model included to enhance understanding?& \\
    \cline{2-3}
    & Are mathematical equations numbered separately and integrated into the text?& \\
    \cline{2-3}
    & Is math mode used for in-line symbols, with bold symbols reserved for vectors?& \\
    \cline{2-3}
    & Are algorithms presented in a specific algorithm mode, emphasizing math symbols and equations rather than long text?& \\
    \cline{2-3}
    & Are tables and figures referred to in the text according to the journal’s formatting guidelines?& \\    \cline{2-3}
    & Is there a balance between text descriptions and visual data representations?& \\    \hline \hline

 \multirow{1}{*}{Results} &   Are the findings of the study presented without interpretation?& \\    \cline{2-3}
    & Is an objective tone maintained throughout the section?& \\    \cline{2-3}
    & Are there results to support every claim made in the section?& \\    \cline{2-3}
    & Is both quantitative and qualitative data included in the report?& \\    \cline{2-3}
    & Are tables, graphs, and figures incorporated to illustrate key findings?
& \\    \cline{2-3}
    & Is the data arranged logically to facilitate comprehension?& \\    \cline{2-3}
    & Are trends and patterns easy to discern without delving into too much detail?& \\    \cline{2-3}
    & Is the data completely focused on what it reveals?& \\    \cline{2-3}
    & Is emphasis placed on clarity to ensure readers can easily understand the findings?& \\    \cline{2-3}
    & Are the findings presented without any interpretation?& \\    \cline{2-3}
    & Is an objective tone maintained throughout the results section?& \\    \cline{2-3}
    & Are there results to support every claim made?& \\    \cline{2-3}
    & Is both quantitative and qualitative data included?& \\    \cline{2-3}
    & Are tables, graphs, and figures used to illustrate key findings?& \\    \cline{2-3}
    & Is the data arranged logically to enhance comprehension?& \\    \cline{2-3}
    & Are trends and patterns easy to discern?& \\    \cline{2-3}
    & Is the data focused on what it reveals?& \\    \cline{2-3}
    & Is clarity emphasized to ensure an easy understanding of the findings?
&\\\hline\hline

    \multirow{1}{*}{Discussion} &  
    Are the results of the study interpreted in relation to the research question?& \\    \cline{2-3}
    & Is a broader interpretation of the findings provided?& \\    \cline{2-3}
    & Are the implications of the findings for the future discussed?& \\    \cline{2-3}
    & Is the significance of the research findings explained?& \\    \cline{2-3}
    & Are the findings compared with existing literature to highlight similarities and differences?& \\    \cline{2-3}
    & Are potential limitations of the study acknowledged?& \\    \cline{2-3}
    & Are future research avenues suggested?& \\    \cline{2-3}
    & Is there a separate section for future research, or is it combined with the conclusion?& \\    \cline{2-3}
    & Are explanations provided for any unexpected results?& \\    \cline{2-3}
    & Is an overview of the research's contribution to the field provided?& \\    \cline{2-3}
    & Are the results interpreted in relation to the research question?& \\    \cline{2-3}
    & Is a broader interpretation of the findings and their implications for the future provided?& \\    \cline{2-3}
    & Are the findings compared with existing literature to highlight similarities and differences?& \\    \cline{2-3}
    & Are potential limitations of the study acknowledged?& \\    \cline{2-3}
    & Are future research avenues suggested, and is there a separate section for them if necessary?& \\    \cline{2-3}
    & Are explanations for unexpected results provided?& \\    \cline{2-3}
    & Is an overview of the research's contribution to the field provided?&\\\hline\hline

    \multirow{1}{*}{Conclusion and Future Work} &     
    Are the main findings and contributions of the research summarized in a clear and concise manner?& \\    \cline{2-3}
    & Is the significance of the research findings explained?& \\    \cline{2-3}
    & Are the practical applications of the findings highlighted?& \\    \cline{2-3}
    & Are the conclusions consistent with the evidence and arguments presented throughout the paper?& \\    \cline{2-3}
    & Is a conclusion written for each objective stated in the introduction?& \\    \cline{2-3}
    & Is the scope and direction of future research specified?& \\    \cline{2-3}
    & Are areas for further research suggested based on the findings?& \\    \cline{2-3}
    & Is the importance and potential impact of the research explained?& \\    \cline{2-3}
    & Is an overview of the paper provided in relation to the introduction and related work?& \\    \cline{2-3}
    & Are the main findings and contributions summarized clearly and concisely?& \\    \cline{2-3}
    & Is the significance of the findings and their practical applications explained?& \\    \cline{2-3}
    & Are the conclusions consistent with the evidence and arguments presented?& \\    \cline{2-3}
    & Is a conclusion provided for each objective stated in the introduction?& \\    \cline{2-3}
    & Is the scope and direction of future research specified?& \\    \cline{2-3}
    & Are areas for further research suggested based on the findings?& \\    \cline{2-3}
    & Is the importance and potential impact of the research explained?& \\    \cline{2-3}
    & Is an overview of the paper provided in relation to the introduction and related work?
    &\\\hline\hline

    \multirow{1}{*}{References} &     
  Are all sources formatted according to a specific citation style (e.g., APA, MLA, Chicago)?& \\    \cline{2-3}
    & Are the authors' names, title, journal name, volume number, page numbers, and publication month/year included for each journal paper?& \\    \cline{2-3}
    & Are the authors' names, title, conference proceedings, page numbers, city, country, and year included for each conference reference?& \\    \cline{2-3}
    & Is “Proc.” used before the conference title?& \\    \cline{2-3}
    & Is emphasis placed on citing recent papers (e.g., last three years) to reflect current research trends?& \\    \cline{2-3}
    & Is preference given to high-quality journal papers over online or ArXiv resources?& \\    \cline{2-3}
    & Is excessive citation of conference papers avoided?& \\    \cline{2-3}
    & Are at least some papers published in the target journal cited?& \\    \cline{2-3}
    & Are titles placed in {{.....}} to indicate acronyms in capital letters?& \\    \cline{2-3}
    & Is care taken to avoid citing any retracted papers?& \\    \cline{2-3}
    & Are all sources formatted according to the specified citation style?& \\    \cline{2-3}
    & Are the necessary details included for each journal paper?& \\    \cline{2-3}
    & Are the necessary details included for each conference reference, with “Proc.” preceding the conference title?& \\    \cline{2-3}
    & Is emphasis placed on citing recent papers to reflect current research trends?& \\    \cline{2-3}
    & Is preference given to high-quality journal papers over online or ArXiv resources?& \\    \cline{2-3}
    & Is excessive citation of conference papers avoided?& \\    \cline{2-3}
    & Are at least some papers published in the target journal cited?& \\    \cline{2-3}
    & Are titles placed in {{.....}} to indicate acronyms in capital letters?& \\    \cline{2-3}
    & Is care taken to avoid citing any retracted papers?  
    &\\\hline
      
\end{longtable}

\end{document}